\documentstyle[prl,twocolumn,aps,eqsecnum,epsf]{revtex}
\begin{document}
\draft
\title{Correlated transport and non-Fermi-liquid behavior in
single-wall carbon nanotubes}
\author{Reinhold Egger$^1$ and Alexander O. Gogolin$^2$}
\address{${}^1$Fakult\"at f\"ur Physik, Albert-Ludwigs-Universit\"at, 
Hermann-Herder-Stra{\ss}e 3, D-79104 Freiburg, Germany\\
${}^2$Department of Mathematics, Imperial College, 180 Queen's Gate,
London SW7 2BZ, United Kingdom}
\date{Date: \today}
\maketitle
\begin{abstract}
We derive the effective low-energy theory for single-wall 
carbon nanotubes including the Coulomb interactions among electrons.
The generic model found here consists of two spin-$\frac12$
fermion chains which are coupled by the interaction.  
We analyze the theory using bosonization, renormalization-group
techniques, and Majorana refermionization.  Several experimentally
relevant consequences of the  breakdown of Fermi liquid
theory observed here are discussed in detail,
e.g., magnetic instabilities, anomalous conductance
laws,  and impurity screening profiles.  
\end{abstract}

\narrowtext

\section{Introduction} \label{sec:intr}

The fascinating electronic and mechanical properties of
carbon nanotubes have 
recently attracted a lot of attention \cite{ebbesen}. 
Nanotubes are tubular nanoscale objects which can be thought of as
graphite sheets wrapped into a cylinder. 
Shortly after their discovery during the carbon-arc
fullerene production \cite{iijima}, theorists have made a
surprising prediction linking structural
to electronic properties \cite{hamada,mintmire,saito,saito2}. 
Specifically, the arrangement of carbon atoms on
the tube surface is determined by the integer indices
$0\leq m \leq n$ of the wrapping superlattice vector 
$\bbox{T}=n\bbox{a}_1+m\bbox{a}_2$, where $\bbox{a}_1$ and 
$\bbox{a}_2$ are the primitive Bravais translation vectors
of the honeycomb lattice. 
Depending on the choice of $n$ and $m$, a $(n,m)$ nanotube
should then be either a metal, a narrow-gap semiconductor,
or an insulator.  This theoretical prediction has been 
amply confirmed 
in recent experiments \cite{dai,ebb,kasumov,thess,direct}.
  
Nanotubes can be fabricated
using either the  carbon-arc method \cite{iijima,prod2}
or a novel laser ablation technique \cite{thess},
where Co- or Ni-doped graphite targets are laser-vaporized. 
The carbon-arc process usually yields multi-wall nanotubes (MWNT) 
composed of several concentric graphite sheets. One can then attach 
metallic leads to such a MWNT and measure, e.g., the
magnetoconductance. 
The MWNT experiment by Langer {\em et al.}\cite{langer}
shows typical signs of a weakly disordered mesoscopic
system, such as universal conductance fluctuations or
weak localization phenomena.  On the other hand, the
laser ablation method allows for the controlled fabrication
of single-wall nanotubes (SWNT).  Rather large 
quantities of metallic $(n,n)$ ``armchair'' nanotubes
with $n=10$ can be obtained 
using this technique \cite{thess}.
In most cases, the SWNTs spontaneously crystallize in
triangular-packed ropes containing $\approx 20$ to 100 SWNTs,
to which one can again attach leads and perform transport 
measurements.  The experimental
results of Fischer {\em et al.} \cite{fischer} have shown a 
linear temperature dependence of the resistance above a
(non-universal) crossover temperature, with an increase of
the resistance at lower temperatures.  In addition, the
experiments of Bockrath {\em et al.} \cite{bockrath} showed
Coulomb charging effects \cite{grabert} in such a rope.

One can in fact also produce single SWNTs using the laser ablation
process.  From a fundamental (and theoretical) point of view, 
a single SWNT is an intriguing system, since 
Coulomb interactions induce a breakdown of
Fermi liquid theory \cite{pines} in any one-dimensional (1D)
metal.  Astonishingly, Tans {\em et al.} \cite{tans} were able
to attach contacts to a single $3\mu$m long  $(10,10)$
armchair SWNT.  The transport measurements of Ref.\cite{tans}
were dominated by charging effects due to 
large contact resistances 
[around 500 k$\Omega$]
between the leads and the SWNT.
By finding a way to decrease the contact resistances one could
 circumvent charging  effects, thereby allowing
one to study the peculiar properties of a 1D conductor.
Nanotubes are potentially much more stable 1D conductors 
than conventional systems such as  
chain molecules \cite{gruner}, 
edge states in the (fractional) quantum Hall
effect \cite{wen1}, or quantum wires in
semiconductor heterostructures \cite{gogolin}. In these systems,
non-Fermi liquid behavior has been notoriously difficult
to establish experimentally.

As has been demonstrated in Ref.\cite{tans},
a metallic SWNT constitutes a perfect experimental realization of 
a 1D conductor.  Interacting 1D electrons exhibit
Luttinger liquid rather than Fermi liquid behavior
characterized by, e.g., the absence of Landau quasi-particles,
spin-charge separation, suppression of the
electron tunneling density of states, and 
interaction-dependent power laws for transport quantities.
For detailed accounts of the physics of Luttinger liquids,
see, e.g., Refs.\cite{voit,book}.
In this paper we describe
the nature of the non-Fermi-liquid state in a SWNT
by deriving and analyzing the
effective low-energy theory of carbon nanotubes. 
This state should be observable as soon as charging effects are 
overcome by lower contact resistances. 
A brief and incomplete account of our results has
been given in Ref.\cite{egger97}.

The role of Coulomb interactions is most pronounced if the
nanotube is metallic. What are the conditions for metallicity? 
A necessary condition arises because
the Fermi vector should obey
the transverse quantization condition $\bbox{T k}=2\pi I$
for an integer $I$. The first Brillouin zone of the honeycomb
lattice is a hexagon, and bandstructure calculations 
\cite{wallace,painter} show that the only gapless points 
are the corner points of this
hexagon. One then finds that there are exactly two independent
Fermi points denoted by $K$ and $K'$, 
with two linearly dispersing 
bands around each of these two points.
Imposing the transverse quantization condition
for the Fermi vector corresponding to $K$ (or $K'$) 
implies $2n+m=3I$. 
If this condition is not fulfilled, the nanotube
exhibits the ``primary'' band gap $\Delta E = 2v/3R\approx$ 1 eV, 
where $v$ is the Fermi velocity and $R$ the  radius.
We mention in passing that a nanotube without primary gap 
has a finite Fermi momentum only if
$n-m=3dI$, where $d$ is the highest common divisor 
of $n$ and $m$ \cite{jishi}. 

Even if the necessary condition $2n+m=3I$ is fulfilled,
the rearrangement of local bonds due to the curvature 
of the cylinder can introduce a ``secondary'' gap,
$\Delta E \approx$ 10 meV, which implies
narrow-gap semiconducting behavior.  
For very small tube diameter, e.g., for a $(6,0)$
tube, due to the strong curvature-induced hybridization 
of $\sigma$ and $\pi$ orbitals,
this effect can be quite pronounced \cite{blase}.
In the case of armchair nanotubes ($n=m$), however, the formation
of a secondary gap is prevented by the high symmetry,
and therefore a $(n,n)$ SWNT stays metallic for all $n$.
The highly symmetric structure of $(n,n)$ [and $(n,0)$ ``zig-zag'']
nanotubes also prohibits chirality, i.e., the
carbon atoms close around the waist of the nanotube. 
In contrast, all other nanotubes show
chiral behavior \cite{russ}.
 
One-dimensional metals are expected to exhibit a
Peierls instability due to the spontaneous formation
of a lattice distortion.  It has been
argued in Refs.\cite{hamada,saito2} that this
instability should not be of any practical importance in nanotubes.
Mean-field estimates of the Peierls transition temperature 
yield very low values ($\approx$ 1 K),
and fluctuations will then tend to wash out the instability even
further.  We therefore neglect coupling to lattice
distortions in the following.  

So far we have discussed the case of a perfectly clean SWNT.
Nevertheless, there are various sources for 
impurities, e.g.,
 structural imperfections of the nanotube (like substitutional
atoms),  charge defects in the substrate, 
topological defects, or twists.  Topological defects are
dislocations (kinks) that locally change the superlattice 
vector $\bbox{T}$ by replacing one of the
hexagons in the graphite network by a pentagon or a heptagon. 
Such a change typically introduces a metal-insulator interface.
By combining  a kink and its antikink (dislocation pair), 
it is possible to form local
metal-insulator-metal junctions \cite{chico}. Transport
is then dominated by tunnel events through this barrier. 
 A less dramatic but probably more
widespread source for backscattering is given by twists of
the nanotube. These twists can originate from the momentary
position of the SWNT when deposited onto the leads, or
from thermal fluctuation modes \cite{kane}.

Generally, especially in the presence of impurities, 
pronounced effects of the Coulomb interaction
on transport in 1D metals can be expected \cite{old}.
As shown in Ref.~\cite{kf}, at very low energy scales
transport is fully blocked by a single arbitrarily weak impurity.
The conductance then 
vanishes as a function of temperature and/or voltage
 with anomalous interaction-dependent power laws.
This effect can be understood in terms of the 
Friedel oscillation \cite{egger95,fabrizio95} 
building up around a scatterer in a 1D metal. 
The oscillatory charge screening cloud displays 
an algebraically slow decay away from the impurity 
and thereby causes an important
additional backscattering contribution \cite{matveev}. 
This crucial effect cannot be captured by linear screening 
\cite{egger97b}.
The dielectric function approach to impurity screening in
SWNTs \cite{liu} is therefore practically useless in a 
determination of the Friedel oscillation.

Below we focus on a single armchair SWNT,
where interaction effects are very pronounced and 
direct experimental tests of the theory are within reach.
However, as long as
the above-mentioned band gaps are negligible,
e.g., for sufficiently high temperatures,
our effective low-energy 
Hamiltonian applies to any SWNT, even a chiral one.
The only change arising for $n\neq m$
affects  the precise value of
the interaction parameters introduced below.
This remarkable circumstance is due to a
 $U(1)$ symmetry of the graphite dispersion relation
present close to the Fermi surface. 
Finally, we mention that
by adding a bulk mass term, the properties of semiconducting 
or insulating SWNTs could be analyzed within the same 
framework \cite{newfoot}.

The structure of this paper is as follows. In Sec.~\ref{sec:eff}
the effective low-energy description of SWNTs is derived.
The resulting fermionic model is bosonized in Sec.~\ref{sec:bos}.
A renormalization group analysis is carried out in Sec.~\ref{sec:rg}
and supplemented by the Majorana refermionization approach described
in Sec.~\ref{sec:maj}.  Experimentally relevant susceptibilities
and correlation functions are discussed in Sec.~\ref{sec:susc}.
In Sec.~\ref{sec:trans}, we study transport through the nanotube,
 followed by a discussion of 
impurity screening in Sec.~\ref{sec:imp}.
Finally, some concluding remarks can be found in Sec.~\ref{sec:conc}.
Throughout this paper, we put $\hbar=1$.

\section{Effective low-energy theory}
\label{sec:eff}

In this section we derive the effective 
low-energy description of a single-wall $(n,n)$ carbon
nanotube.  Our description applies at energy scales where the
generic linear dispersion relation of a SWNT depicted
in Fig.~\ref{fig1} is valid.  For a
$(10,10)$ nanotube, the theory therefore holds at room 
temperature and below.  A similar approach has been devised
previously for the uncorrelated case by 
Kane and Mele \cite{kane}, building on earlier work
\cite{divincenzo} for graphite intercalation
compounds.  Here we extend their theory and include the
Coulomb interactions between the electrons.
Previously, this problem has 
only been investigated using the perturbative
renormalization group (RG) for a weak short-range (Hubbard) 
interaction \cite{krotov,balents}.  However, there is no external
screening of the Coulomb interaction in the SWNT experiments of
Ref.\cite{tans}, and one has to take into account the
long-ranged character of the
Coulomb interaction potential (see also Ref.\cite{kanenew}). 
Since one might employ suitable screening backgates in future
experiments, the general (short- or long-ranged) case of an
arbitrary interaction potential is treated here.
The method used below allows us
to get insight into the physics of the strong-coupling regime
emerging at low temperatures. 

\subsection{Low-energy expansion}

The remarkable electronic properties of carbon nanotubes are
due to the special bandstructure of the $\pi$ electrons in
graphite. The simplest bandstructure calculation starts 
from a nearest-neighbor
tight-binding Hamiltonian on the honeycomb lattice, 
which can be straightforwardly 
diagonalized \cite{wallace}.  Remarkably, the only gapless
points of the resulting dispersion relation are the corner points
of the hexagonal first Brillouin zone. Hence
there are only two linearly independent Fermi points
$\alpha \bbox{K}$ with $\alpha=\pm$ 
instead of a continuous Fermi surface. 
Up to energy scales $\leq 1$ eV, 
the dispersion relation around the Fermi points 
is, to a very good approximation, linear,
\begin{equation}\label{egraph}
  E_\pm(\bbox{q}=\bbox{k}-\bbox{K}) = \pm v |\bbox{q}| \;,
\end{equation}
with the same relation around the other Fermi point.
The $+$ ($-$) sign corresponds to the conduction (valence)
band, respectively.
Obviously, close to the Fermi surface a $U(1)$ symmetry
holds, since the direction of $\bbox{q}$ does
not matter in Eq.~(\ref{egraph}). 

Since the basis of the honeycomb lattice contains two atoms,
there are two sublattices $p=\pm$ shifted by the vector
$\bbox{d}=(0,d)$, with the nearest-neighbor C-C distance
$d=a/\sqrt{3}=1.42${\AA} (here $a$ denotes
the honeycomb lattice constant).
As a consequence, there are  
two degenerate Bloch states 
\begin{equation}\label{bloch}
\varphi_{p\alpha}(\bbox{r}) = 
(2\pi R)^{-1/2} \exp( -i\alpha \bbox{K r} ) 
\end{equation}
at each Fermi point $\alpha=\pm$,
where  $\bbox{r}=(x,y)$  lives
on the  sublattice $p$ under consideration.
In Eq.~(\ref{bloch}), we have already anticipated
the correct normalization for nanotubes.
We follow Ref.\cite{divincenzo}
and choose the Bloch functions separately on each
sublattice such that they vanish on the other.
One can then expand the electron operator
in terms of these Bloch waves ($\sigma=\pm$ is
the spin index), 
\begin{equation}\label{expa0}
\Psi_\sigma(x,y) = \sum_{p\alpha} \varphi_{p\alpha}(x,y) 
\, F_{p\alpha\sigma} (x,y) \;,
\end{equation} 
which introduces
slowly varying operators $F_{p\alpha\sigma}(x,y)$.
The resulting
second-quantized effective low-energy theory of
a graphite sheet is given by 
the 2D massless Dirac Hamiltonian \cite{divincenzo},
\begin{equation}\label{2d}
H_G = -i v \sum_{pp'\alpha\sigma} \int d\bbox{r} \, 
F_{p\alpha\sigma}^\dagger
  (\bbox{\tau \nabla})^{}_{pp'}
 F_{p'\alpha\sigma}^{} \;,
\end{equation}
in accordance with the spectrum (\ref{egraph}).
Here  $\bbox{\tau}=(\tau_x,\tau_y)$
are standard Pauli matrices.

\begin{figure}
\epsfysize=10cm
\epsffile{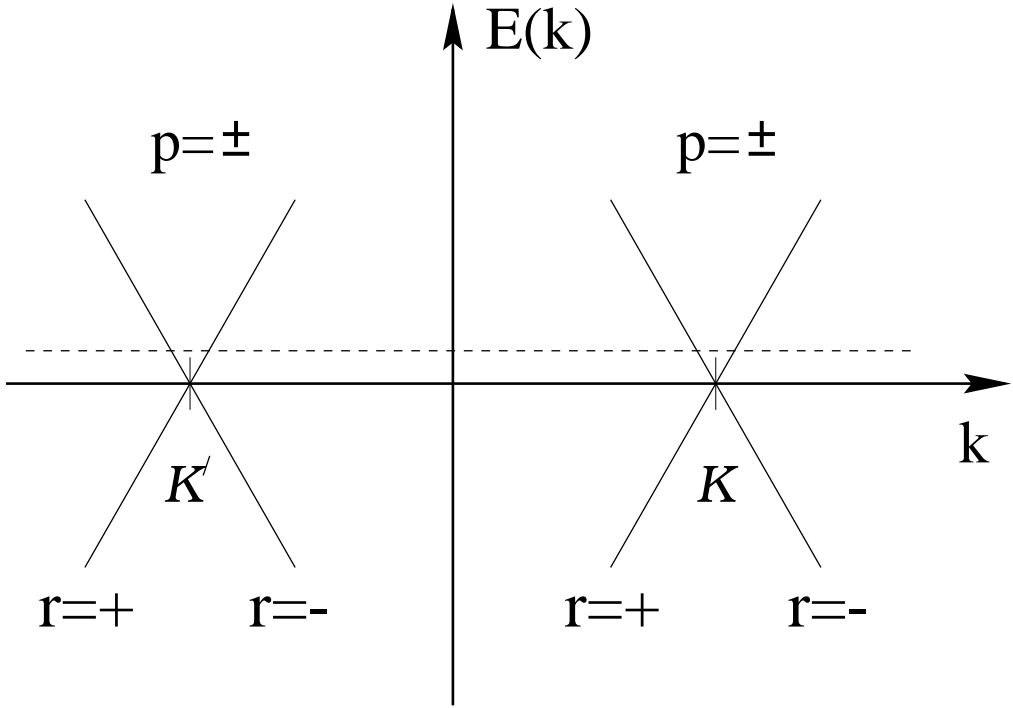}
\caption[]{\label{fig1} Schematic bandstructure of a metallic SWNT.
A right- and left-moving branch ($r=\pm$) is found near each of
 the two Fermi points $\alpha=\pm$ corresponding to
$K$ and $K'$, respectively. Right- and left-movers arise as
linear combinations of the sublattices $p=\pm$. The Fermi energy
(dashed line) can be tuned by an external gate.}
\end{figure}

Wrapping the graphite sheet 
leads to the generic bandstructure of a metallic SWNT
shown in Fig.~\ref{fig1}.  
For a $(n,n)$ armchair SWNT, 
the Fermi vector is $\bbox{K}= (k_F,0)$
with $k_F=4\pi/3a$.
We take the $x$-axis  along the tube direction, and 
the circumferential variable is $0< y <2\pi R$. 
The armchair SWNT radius is
$R=\sqrt{3} na/2\pi$, which yields $R=1.38$ nm for a
 $(10,10)$ nanotube.
Quantization of transverse motion now allows for a
contribution $\exp(i m y/R)$ to the wavefunction. 
However,  excitation of angular momentum states
other than $m=0$ costs the energy $\approx 10$ eV$/n$. 
In an effective low-energy theory, we may thus omit all transport
bands except $m=0$ and hence arrive at a 1D situation with $k_y=0$.
Since also $K_y=0$, the Bloch states (\ref{bloch})
do not depend on the 
circumferential coordinate, and the corresponding
wavefunctions are indeed stiff around the waist of the
nanotube.   

Instead of the
low-energy expansion (\ref{expa0}),  the
electron operator is now written as \cite{kane}
\begin{equation}\label{expa}
\Psi_\sigma(x,y) = \sum_{p\alpha} \varphi_{p\alpha}(x,y) 
\,\psi_{p\alpha\sigma} (x) \;,
\end{equation} 
which introduces
 1D fermion operators $\psi_{p\alpha\sigma}(x)$ depending 
only on the $x$ coordinate.  They correspond to the $F_{p\alpha\sigma}$
for the graphite sheet.
Neglecting Coulomb interactions for the moment, 
the Hamiltonian can be read off from Fig.~\ref{fig1}, 
\begin{equation} \label{h0}
H_0= -v \sum_{p\alpha\sigma} p \int dx \;\psi_{p\alpha\sigma}^\dagger
\partial_x \psi^{}_{-p\alpha\sigma} \;.
\end{equation}
For this expression, we have chosen a preferred direction 
in Eq.~(\ref{2d}).  However, due to the mentioned $U(1)$ invariance,
we can choose any direction without affecting the
resulting low-energy Hamiltonian [we have explicitly checked
that this holds even in the presence of interactions].
Switching from the sublattice
($p=\pm$) description to the right- and left-movers ($r=\pm$) 
indicated in Fig.~\ref{fig1} implies two copies ($\alpha=\pm$)
of  massless 1D Dirac Hamiltonians. 
By using a suitable  gate, one can experimentally tune the
average charge density on the nanotube and hence 
adjust the Fermi energy.  
In contrast to the half-filled band 
($E_F=0$) encountered in 2D graphite sheets, one is
normally off half-filling in SWNT experiments.
Typically, the Fermi energy is displaced by about 300 meV
in the experiment of Ref.~\cite{tans}.
Finally, to describe semi-conducting or 
insulating SWNTs with a band gap $\Delta E$, 
a bulk mass term should be added \cite{newfoot},
\begin{equation} \label{mass}
H' = \frac{\Delta E}{2}
 \int dx \sum_{p\alpha \sigma} \psi^\dagger_{p\alpha\sigma}
\psi^{}_{-p\alpha\sigma} \;.
\end{equation}
We will put $\Delta E=0$ in the following, as is appropriate
for armchair nanotubes.

\subsection{Coulomb interaction}

Let us now examine Coulomb interactions mediated by an arbitrary
potential $U(\bbox{r}-\bbox{r}')$. The interaction
is described by the Hamiltonian
\begin{eqnarray}\label{int0}
H_I &=& \frac12 \sum_{\sigma\sigma'}\int d\bbox{r} \int
d\bbox{r}' \, \Psi^\dagger_\sigma(\bbox{r}) \Psi^\dagger_{\sigma'}
(\bbox{r}') \\
\nonumber && \quad \times \; U(\bbox{r}-\bbox{r}') \Psi^{}_{\sigma'}(\bbox{r}')
\Psi^{}_\sigma(\bbox{r}) \;.
\end{eqnarray}
Herein we have assumed that the interaction 
is not sensitive to the electron
spin $\sigma=\pm$. 
Bound electrons  can  be incorporated in 
terms of a background dielectric constant $\kappa$,
but free charge carriers in nearby gates could lead to
an effectively short-ranged potential. 

For the experiments of Ref.\cite{tans}, 
one has an externally unscreened Coulomb interaction,
\begin{equation}\label{unsc}
U(\bbox{r}-\bbox{r}') = \frac{e^2/ \kappa}
{\sqrt{ (x-x')^2 + 4R^2 \sin^2[(y-y')/2R] + a_z^2 }}\;,
\end{equation}
where $a_z\simeq 3 a_B\approx a$
[with the Bohr radius $a_B=\hbar^2/me^2=0.529${\AA}]
denotes the average distance between a
$2p_z$ electron and the nucleus, i.e.,
the ``thickness'' of the graphite sheet.
The dielectric constant in Eq.~(\ref{unsc}) can be
estimated from the following
elementary consideration. In a strictly 1D system of length $L$ with
interaction potential $u(x-x')=e^2/\kappa|x-x'|$,
 the charging energy is 
\begin{eqnarray*}
E_c &=& \frac{1}{2L^2} \int_0^L dx dx'\, u(x-x')\\
&\simeq&  e^2 \ln(L/R)/\kappa L\;.
\end{eqnarray*}
The experimental value $E_c = 2.6$ meV \cite{tans} 
for $L=3\mu$m leads to $\kappa \approx 1.4$. 
The theoretical estimate $\kappa\approx 2.4$ 
for bulk graphite \cite{taft} is of the same order of magnitude. 
Our estimate also includes the effect of the insulating substrate.

The interaction (\ref{int0}) can now be reduced to a 
1D interaction by inserting the expansion (\ref{expa}) 
for the electron field operator. The result is
\begin{eqnarray}\label{general}
H_I&= &\frac12 \sum_{pp'\sigma\sigma'} 
\sum_{\alpha_1\alpha_2\alpha_3\alpha_4}
 \int dx dx'\; V^{pp'}_{\{\alpha_i\}}(x-x') \\ \nonumber
&\times& 
\psi^\dagger_{p\alpha_1\sigma}(x) \psi^\dagger_{p'\alpha_2\sigma'}
(x') \psi^{}_{p'\alpha_3\sigma'}(x') \psi^{}_{p\alpha_4\sigma}(x) 
\;,
\end{eqnarray} 
with the 1D interaction potentials
\begin{eqnarray}\label{intpot}
&& V^{pp'}_{\{\alpha_i\}}(x-x') = \int_0^{2\pi R} dy dy' \; 
\varphi^{\ast}_{p\alpha_1}(x,y)
\varphi^{\ast}_{p'\alpha_2}(x',y') \\ &\times& \nonumber
U(x-x',y-y'+p d \delta_{p,-p'} )\; 
\varphi^{}_{p'\alpha_3}(x',y') \varphi^{}_{p\alpha_4}(x,y) \;.
\end{eqnarray} 
These potentials only depend on $x-x'$ and on the 1D
fermion quantum numbers.
For interactions involving different sublattices for $\bbox{r}$
 and $\bbox{r}'$ in Eq.~(\ref{int0}), i.e., $p\neq p'$, 
one needs to take into account the shift vector $\bbox{d}=(0,d)$,
see Eq.~(\ref{intpot}).  

\begin{figure}
\epsfysize=12cm
\epsffile{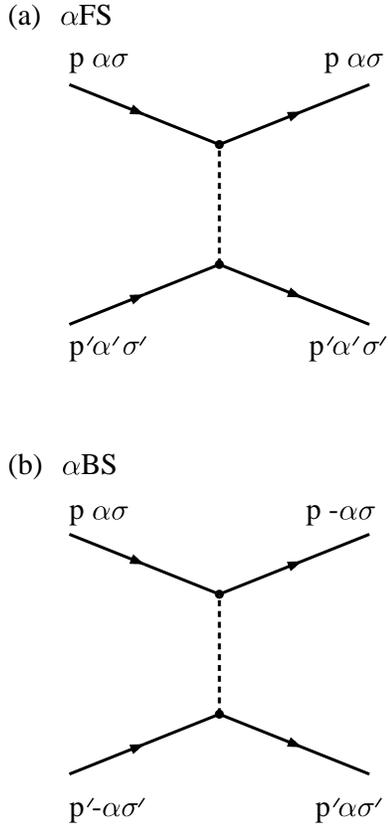}
\caption[]{\label{fig2} Allowed interaction processes away from 
half-filling. (a) Forward scattering ($\alpha$FS). The incoming
1D fermions do not change their quantum numbers $p,\alpha,\sigma$
during the scattering event.
(b) Backward scattering ($\alpha$BS). One 1D fermion is scattered
from Fermi point $\alpha$ to the opposite Fermi point $-\alpha$,
with the other being scattered from $-\alpha$ to $\alpha$. 
Straight lines denote 1D fermion propagators, dashed lines the
effective 1D Coulomb interaction.}
\end{figure}

To simplify the resulting 1D interaction (\ref{general}), 
we now exploit momentum conservation.
Since  an additional gate voltage tunes the
average electron density on the SWNT, we 
assume that no lattice instabilities due
to  Umklapp scattering are present, i.e.,
one stays off half-filling.  This assumption
does not limit the applicability of our approach, 
but simplifies matters considerably.
For a recent study of Umklapp processes in SWNTs, see
Ref.\cite{balents}. We then employ the
Fermi point quantum numbers $\alpha_i$ to classify the 
allowed processes, see Fig.~\ref{fig2}.
First, there are ``forward scattering''
($\alpha$FS) processes,
where $\alpha_1=\alpha_4$ and $\alpha_2=\alpha_3$.
Second, we have ``backscattering'' ($\alpha$BS), with $\alpha_1=-\alpha_2=
\alpha_3=-\alpha_4$.
Notice that the above classification differs from the
conventional one for the standard two-chain problem \cite{book}.
The latter is based on the right- and left-moving indices
($r=\pm$ or $p=\pm$) rather than on the different Fermi points
of the nanotube ($\alpha=\pm$).

\subsection{Forward scattering}

Let us start with $\alpha$FS, where $\alpha_1=\alpha_4$
and $\alpha_2=\alpha_3$.  The corresponding diagram is 
shown in Fig.~\ref{fig2}. We first define the potential
\begin{equation}\label{v0}
V_0(x-x')= \int_0^{2\pi R} \frac{dy}{2\pi R}
\int_0^{2\pi R} \frac{dy'}{2\pi R}\;  U(\bbox{r}-\bbox{r}')\; .
\end{equation}
This gives from Eqs.~(\ref{intpot}) and (\ref{bloch})
 the $\{\alpha_i\}$-independent
 forward scattering  interaction potential
\begin{equation}
V^{pp'}_{\alpha{\rm FS}}(x) = V_0(x)+\delta_{p,-p'} \delta V_p(x)\;,
\end{equation}
with the correction term
\begin{equation}\label{deltav}
 \delta V_p(x)  =  \int_0^{2\pi R} \frac{dy dy'}{(2\pi R)^2}
 [U(x,y-y'+p d)- U(x,y-y') ] \;,
\end{equation}
which is only present if $\bbox{r}$ and $\bbox{r}'$ are
located on different sublattices.  Thereby important
 information about the discrete nature of the graphite 
network is retained despite
the low-energy continuum approximation employed in our
formulation. The correction $\delta V_p$
is a direct measure of the 
difference between intra- and inter-sublattice
interactions.
Because of the periodicity of the $y$-integrals, Taylor expanding 
Eq.~(\ref{deltav}) in powers of $d$ implies that the correction 
$\delta V_p(x)=0$. 
Since  $V_0(x)$ treats both sublattices on equal footing,
 the resulting $\alpha$FS interaction couples only the
total electron densities, 
\begin{equation}\label{fs0}
H_{\alpha{\rm FS}}^{(0)} = \frac12 \int dx dx' \, 
\rho(x) V_0(x-x') \rho(x') \;,
\end{equation}
where the 1D density is defined as
\begin{equation}\label{cd}
\rho(x) = \sum_{p\alpha\sigma} \psi^\dagger_{p\alpha\sigma}(x)
\psi^{}_{p\alpha\sigma}(x) \;.
\end{equation} 
This density does not contain ``fast''
 terms arising from a mixture of
the different ($p\alpha\sigma$)-type fermions. Such terms turn out to be
crucial for an understanding of the Friedel oscillation in
Sec.\ref{sec:imp}.

For the unscreened Coulomb interaction  (\ref{unsc}), the
 potential (\ref{v0}) becomes 
\begin{equation}\label{vv0}
V_0(x)=\frac{2e^2} {\kappa\pi\sqrt{a_z^2+x^2+4R^2}} 
\,K \left(\frac{2R}{\sqrt{a_z^2+x^2+4R^2}}\right)\;,
\end{equation}
with the complete elliptic 
integral $K(z)$ of the first kind \cite{gradsteyn}.
For $x\gg R$, one finds $V_0(x)\sim 1/x$ again.
The Fourier transform $\widetilde{V}_0(k)=\widetilde{V}_0(-k)$ 
can be expressed in terms of the modified Bessel function
$K_0(z)$ \cite{abramowitz},
\begin{equation}\label{v0k}
\widetilde{V}_0(k) = \frac{4e^2}{\kappa \pi} 
\int_0^{\pi/2} d \varphi\, K_0 \left (
k\sqrt{a_z^2+4R^2 \sin^2 \varphi}\right)\;.
\end{equation}
Using asymptotic properties of  $K_0(z)$, 
we then find the long-wavelength form valid
for $kR\ll 1$, 
\begin{equation}\label{un2}
\widetilde{V}_0(k) =  \frac{2 e^2}{\kappa} ( |\ln kR| + c_0 )  \;,
\end{equation}
where $c_0=-\gamma+(\pi/2) \ln 2 \simeq 0.51$
with Euler's constant $\gamma$.
The logarithmic singularity at small wavevectors 
is a direct consequence of the long-range tail of $V_0(x)$.

If one starts instead
 from an on-site
Hubbard interaction \cite{krotov,balents},
\begin{equation}\label{hubb1}
U(\bbox{r}-\bbox{r}') = U \delta_{\sigma,-\sigma'} \delta_{pp'}
\delta(\bbox{r} -\bbox{r}') \;,
\end{equation}
one finds from Eq.~(\ref{v0})
 the 1D forward scattering potential 
\begin{equation}
V_0(x)= \frac{U}{2\pi R} \delta_{\sigma,-\sigma'} \delta(x-x')\;.
\end{equation}
The characteristic $1/R$ scaling of the effective 1D interaction
has also been found in Ref.\cite{balents}. 

\begin{figure}
\epsfysize=10cm
\epsffile{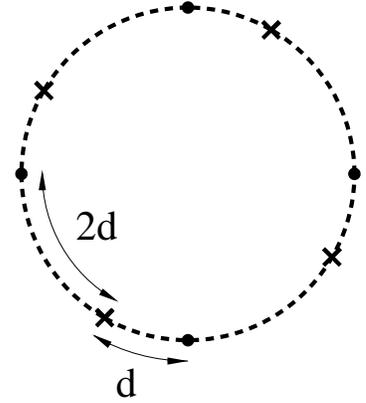}
\caption[]{\label{fig3}  Microscopic arrangement
of carbon atoms around the waist of an armchair
SWNT (shown for $n=4$).  Circles and crosses denote
the two distinct sublattices $p=\pm$,
and $d=a/\sqrt{3}$.}
\end{figure}

For $|x-x'|\gg a$,  our  continuum calculation leading to
 $\delta V_p(x-x')=0$ is  justified. However, 
for $|x-x'|\leq a$, one has to be more careful.
Here an additional $\alpha$FS term beyond Eq.~(\ref{fs0}) arises
due to the hard core of the Coulomb interaction.  To study this term,
we put $x=x'$ and evaluate $\delta V_p(0)$ directly on the wrapped 
graphite network shown in Fig.~\ref{fig3}.
For simplicity, we again consider an armchair SWNT,
albeit the same result [see Eq.~(\ref{fs1}) below]
is found for other nanotube geometries.
We now write
\begin{equation} \label{intermed}
\delta V_p (x-x') \simeq - 2 f_p \delta(x-x')\;,
\end{equation}
and since $f_p=f_{-p}=f$ [see below], we obtain
the additional $\alpha$FS contribution in the form
\begin{equation}\label{fs1}
H^{(1)}_{\alpha{\rm FS}}= 
- f \int dx\sum_{p\alpha\alpha'\sigma\sigma'}
\psi^\dagger_{p\alpha\sigma}\psi^\dagger_{-p\alpha'\sigma'}
\psi^{}_{-p\alpha'\sigma'} \psi^{}_{p\alpha\sigma} \;.
\end{equation}
The coupling constant $f>0$ depends on the nanotube geometry 
only and is evaluated for the $(n,n)$ case next. 

To calculate $\delta V_p(0)$ on the wrapped graphite lattice,
we start from the microscopic
arrangement of carbon atoms around the waist of the
armchair SWNT, see Fig.~\ref{fig3}. On the length scale
$|y-y'|\leq 2\pi R$, it is always justified to use the externally
unscreened Coulomb interaction potential (\ref{unsc}).
Denoting the locations of carbon atoms on sublattice $p=+$
(the circles in Fig.~\ref{fig3}) as $y_k=2\pi (k/n) R=3kd$ with
$k=1,\ldots,n$, and discretizing Eq.~(\ref{deltav}), we obtain
from Eq.~(\ref{intermed})
\begin{eqnarray*}
2 f_p/a &=& (3d/2\pi R)^2  \sum_{l,k=1}^{n} \{
U(0,y_l-y_k) \\ && \quad - U(0,y_l-y_k+pd) \} \;.
\end{eqnarray*}
The summation extends over the $p=+$ sites only,
and inserting Eq.~(\ref{unsc}) yields with 
 $R=\sqrt{3} n a/2\pi$, 
\begin{eqnarray} \label{sumint}
2f_p/a &=& \frac{e^2}{2n^2 \kappa R} \sum_{l,k}
\Bigl \{
 \frac{1}{\sqrt{\sin^2[(l-k)\pi/n] + (a_z/2R)^2}} \\
\nonumber && \quad - 
 \frac{1}{\sqrt{\sin^2[(l-k+p/3)\pi/n] + (a_z/2R)^2}} 
\Bigr \} \;.
\end{eqnarray}
The singular contributions are picked up
 from $l=k$ and $l=n-k$ and
yield the $p$-independent contribution
\begin{equation} \label{gam}
f/a = \gamma^{}_f \, e^2/R\;,
\end{equation}
with the
order-of-magnitude estimate 
\[
\gamma^{}_f =  \frac{\sqrt{3} a}{2\pi\kappa a_z} \left[ 1- \frac{1}
{\sqrt{1+ a^2/3 a_z^2}} \right] \approx 0.05 \;.
\]
It can be checked easily that
the remaining terms in the summation (\ref{sumint}) 
are negligible against the singular contribution leading to
Eq.~(\ref{gam}).  The prefactor $\gamma_f$
is independent of $n$ such that $f\sim 1/n$. 
Parenthetically, in the
 language of the Hubbard-like models employed in
Refs.~\cite{krotov,balents}, 
we have $f/a= U-V$, where $U\approx e^2/R$
is the on-site and $V$ the nearest-neighbor 
Coulomb interaction. 
According to Eq.~(\ref{gam}), this difference is 
small compared to $U$.  Using a strictly on-site 
Hubbard model $(V=0)$ is therefore never
realistic, even in the presence of close-by screening
gates.

\subsection{Backward scattering}

Let us now discuss the $\alpha$BS contributions depicted
in Fig.~\ref{fig2}. For backscattering processes,
we have $\alpha_1=-\alpha_2=\alpha_3=-\alpha_4=\alpha$
in Eq.~(\ref{general}), and from Eq.~(\ref{intpot})
the corresponding 1D interaction potential for $\alpha$BS processes reads
\begin{equation}\label{bspot}
V_\alpha^{pp'}(x-x') = e^{2i\alpha k_F (x-x')}\;
V_{\alpha{\rm FS}}^{pp'}(x-x')  \;.
\end{equation}
Because of the rapidly oscillating phase factor, 
the only non-vanishing  contribution to $H_{\alpha{\rm BS}}$
comes from $|x-x'|\leq a$, i.e., we can effectively take a local 
interaction.  
According to Eq.~(\ref{general}), the $\alpha$BS contribution
must then be of the general form
\[
H_{\alpha{\rm BS}}= \frac{1}{2} \int dx\sum_{pp'\alpha\sigma\sigma'}
(b+pp' b')
\psi^\dagger_{p\alpha\sigma}\psi^\dagger_{p'-\alpha\sigma'}
\psi^{}_{p'\alpha\sigma'} \psi^{}_{p-\alpha\sigma} \;.
\]
The term corresponding to $b'$ is irrelevant \cite{odintsov}
and is omitted in what follows. 
The effective coupling constant $b>0$ is then determined by 
the Fourier transform of $V_0(x)$ at $k=2k_F$,
\begin{equation}\label{bb}
b = \widetilde{V}_0(2k_F) \;,
\end{equation}
and the backscattering contribution is
\begin{equation} \label{bs}
H_{\alpha{\rm BS}}= \frac{b}{2} \int dx\sum_{pp'\alpha\sigma\sigma'}
\psi^\dagger_{p\alpha\sigma}\psi^\dagger_{p'-\alpha\sigma'}
\psi^{}_{p'\alpha\sigma'} \psi^{}_{p-\alpha\sigma} \;.
\end{equation}
For the unscreened interaction (\ref{unsc}), 
using $2k_F R= 4n/\sqrt{3} \gg 1$ and asymptotic properties
of the Bessel function, we obtain from Eq.~(\ref{v0k}) 
\begin{equation}\label{bbb}
b/a=\gamma^{}_b\, e^2/R \;,
\end{equation}
with the order-of-magnitude estimate 
\[
\gamma^{}_b \approx  \frac{3}{2\pi^2 \kappa} \approx 0.1 \;.
\]
The prefactor $\gamma_b$ is independent of $n$,
and thus $b\sim 1/n$.
This calculation for unscreened interactions predicts 
$b\approx f$. If the Coulomb interaction
is externally screened, however,
a qualitatively different situation can arise.
Now $f$ is still given by 
Eq.~(\ref{gam}), but $b$ can 
become significantly larger, see Eq.~(\ref{bb}).
Therefore it is possible to have $b\gg f$ in the presence
of screening gates.  

\section{Bosonization}
\label{sec:bos}

According to Sec.~\ref{sec:eff},
the low-energy theory of armchair SWNTs away
from lattice commensurabilities is 
described by the Hamiltonian
\begin{equation} \label{ham1}
H = H_0 + H_{\alpha{\rm FS}}^{(0)} +
H_{\alpha{\rm FS}}^{(1)} + H_{\alpha{\rm BS}}^{} \;.
\end{equation}
This model is equivalent to
two  spin-$\frac12$ fermion chains coupled in a rather special
way by the interactions, but without 
interchain single-particle hopping.
As described below, it is in that respect that
our theory differs from 
the standard two-chain problem of coupled
Luttinger liquids investigated, e.g., in 
Refs.\cite{kusmartsev,fabrizio,finkel,rice,schulz,kopietz,giamarchi}.
The standard two-chain
 model is usually derived by coupling two Hubbard chains
via a transverse hopping matrix element \cite{noack},
and has been used to study, e.g.,
the stability of  Luttinger liquid 
behavior with respect to interchain coupling
 \cite{anderson,wen2,castellani}. In contrast,
the modified model (\ref{ham1}) describes
the properties of a SWNT.  Our solution of this model
proceeds in fact quite similar
to the standard two-chain case.
In particular, we now follow the
bosonization route taken in
Refs.\cite{fabrizio,finkel,rice} and especially in
Ref.\cite{schulz}.  For a general review of 
bosonization, see Ref.\cite{book}.

In order to proceed,
we need to bring the non-interacting
Hamiltonian (\ref{h0}) into a standard form
of the 1D Dirac model.  This is accomplished by
switching to right- and left-movers
($r=\pm$) which are linear combinations of the sublattice
states $p=\pm$. In particular, we have 
\begin{equation} \label{rotate}
\psi_{p\alpha\sigma}(x) = \sum_r \widetilde{U}_{pr} 
\widetilde{\psi}_{r\alpha\sigma}(x) \;,
\end{equation}
where the unitary operator $\widetilde{U}$ fulfills  
$\widetilde{U}^\dagger \sigma_y \widetilde{U} = \sigma_z$.
Since the indices of $\widetilde{U}$ are $p=\pm$ and $r=\pm$,
the $2\times 2$ matrix representation
$\widetilde{U} =  (1/\sqrt{2}) [(1,i),(1,-i)]$ is easily found. 
The non-interacting Hamiltonian (\ref{h0})
then reads
\begin{equation}\label{h02}
H_0= -i v
 \sum_{r\alpha\sigma} r \int dx \;
\widetilde{\psi}_{r\alpha\sigma}^\dagger
\partial_x \widetilde{\psi}^{}_{r\alpha\sigma} \;.
\end{equation}
Furthermore, the 1D density operator 
$\rho_{p\alpha\sigma}(x)$ entering Eq.~(\ref{cd})
becomes
\begin{equation} \label{dens}
\rho_{p\alpha\sigma} = 
\psi^\dagger_{p\alpha\sigma}
\psi^{}_{p\alpha\sigma}
= \frac12 \sum_{r=\pm} \left(\widetilde{\rho}_{r\alpha\sigma}
+p 
\widetilde{\psi}^\dagger_{r\alpha\sigma}
\widetilde{\psi}^{}_{-r\alpha\sigma} \right) \;,
\end{equation}
where $\widetilde{\rho}_{r\alpha\sigma}(x) =
\widetilde{\psi}^\dagger_{r\alpha\sigma}(x)
\widetilde{\psi}^{}_{r\alpha\sigma}(x)$.
The 1D density  
defined in Eq.~(\ref{cd}) is therefore equivalently 
expressed as 
\begin{equation}\label{tode}
\rho(x)=\sum_{r\alpha\sigma}
\widetilde{\psi}^\dagger_{r\alpha\sigma}(x)
\widetilde{\psi}^{}_{r\alpha\sigma}(x)\;.
\end{equation}

In this representation, it is convenient to
apply the bosonization formula \cite{book}, 
\begin{equation}\label{boson}
\widetilde{\psi}_{r\alpha \sigma}(x) =
\frac{\eta_{r\alpha \sigma}}{\sqrt{2\pi a}}
\exp\left\{ iq_F r x+ik_F \alpha x + i \varphi_{r\alpha\sigma} \right\} \;.
\end{equation}
For simplicity,  we have incorporated the spatial dependence due to
the Bloch functions (\ref{bloch}) into the 1D fermion operator.
The chiral bosonic phase fields
obey the commutator algebra
\begin{equation}\label{alg1}
 \left[ \varphi_{r\alpha\sigma}(x), \varphi_{r'\alpha'\sigma'}(x')
\right]_- = -i\pi r \delta_{rr'}\delta_{\alpha \alpha'}
\delta_{\sigma\sigma'}\; {\rm sgn}(x-x') \;.
\end{equation}
From these relations, the fermion operator 
$\widetilde{\psi}_{r\alpha\sigma}(x)$ has indeed
the correct anticommutator algebra on a given
branch $(r\alpha\sigma)$.  To enforce anticommutator
relations between different branches, we need the
Majorana fermions (Klein factors) $\eta^{}_{r\alpha\sigma}
=\eta^\dagger_{r\alpha \sigma}$.  They obey
\begin{equation}\label{majalg}
\left [ \eta_{r\alpha\sigma}, 
\eta_{r'\alpha'\sigma'} \right ]_+ = 2 \delta_{rr'}
\delta_{\alpha\alpha'} \delta_{\sigma \sigma'} \;.
\end{equation}
We note that the usual zero modes \cite{book} 
have to be  incorporated in the  fields
$\varphi_{r\alpha\sigma}$ as they are {\em not}\, contained in the
Majorana fermions.  However, they matter only if one is
interested in finite-size effects.
Finally, from Eq.~(\ref{boson}),  the density
 $\widetilde{\rho}_{r\alpha \sigma}$
now takes the form
\[
 \widetilde{\rho}_{r\alpha \sigma} (x)= \frac{q_F}{2\pi}
+ \frac{r}{2\pi} \partial_x \varphi_{r\alpha\sigma}(x)
\;.
\]
The wave vector $q_F$, which must be carefully distinguished
from the Fermi vector $k_F$, is related to the band
filling.  By varying a suitable gate voltage \cite{tans},
one can easily adjust the Fermi energy and hence
the band filling (see Fig.~\ref{fig1}).  Relative
to the unbiased half-filled case $E_F=0$,  an
average excess density $\delta\rho$ gives rise
to a non-zero $q_F = \pi \delta \rho/4$ and hence
$E_F=v q_F$.  We consider $|q_F|\ll k_F$ in the following,
since otherwise the low-energy continuum approximation
underlying our approach might break down. 

At this stage, it is natural to introduce the
standard linear combinations $\theta_{j\delta}(x)$
and their dual fields $\phi_{j\delta}(x)$ subject to
the algebra
\begin{equation} \label{alg2}
[\phi_{j\delta}(x),\theta_{j'\delta'}(x')] 
= -(i/2) \delta_{jj'}\delta_{\delta\delta'}\, {\rm sgn}(x-x') \;.
\end{equation}
The bosonic   phase fields $\theta_{j\delta}(x)$
 for  the total and relative 
($\delta=\pm$) charge ($j=c$) and
spin ($j=s$) channels are constructed  as
\begin{eqnarray*}
\theta_{c+} &=& \frac{1}{4\sqrt{\pi}} \sum_{r\alpha\sigma} 
r\, \varphi_{r\alpha\sigma} \;, \\
\theta_{c-} &=& \frac{1}{4\sqrt{\pi}} \sum_{r\alpha\sigma} 
r\alpha \,\varphi_{r\alpha\sigma} \;, \\
\theta_{s+} &=& \frac{1}{4\sqrt{\pi}} \sum_{r\alpha\sigma} 
r\sigma\, \varphi_{r\alpha\sigma} \;, \\
\theta_{s-} &=& \frac{1}{4\sqrt{\pi}} \sum_{r\alpha\sigma} 
r\alpha\sigma \, \varphi_{r\alpha\sigma} \;.
\end{eqnarray*}
Their dual fields $\phi_{j\delta}$ are defined similarly by
omitting the $r$ factor in these relations.
The back-transformation reads 
\begin{eqnarray}\label{bos2}
\varphi_{r\alpha\sigma} &=& \frac{\sqrt{\pi}}{2} \Bigl \{
\phi_{c+} + r \,\theta_{c+} +
\alpha \,\phi_{c-} + r\alpha\, \theta_{c-} 
\\ \nonumber & +&
\sigma\, \phi_{s+} + r\sigma\, \theta_{s+} +
\alpha \sigma\, \phi_{s-} + r\alpha\sigma\, \theta_{s-} \Bigr \} \;,
\end{eqnarray}
such that Eq.~(\ref{alg1}) is recovered from Eq.~(\ref{alg2}).
The total electron density relative to
the half-filled situation reads from Eq.~(\ref{tode})
\begin{equation}\label{tode2}
\rho(x) = 4q_F/\pi+ \frac{2}{\sqrt{\pi}} \partial_x \theta_{c+}(x)\;,
\end{equation}
and the continuity equation then yields the current,
\begin{equation}\label{curr}
I = \frac{2e}{\sqrt{\pi}} \partial_t \theta_{c+}(x,t)\;,
\end{equation}
which can be evaluated, say, at $x=0$.

In order to arrive at the bosonized form of the model
(\ref{ham1}), we need to specify the Majorana fermions
$\eta_{r\alpha\sigma}$ in Eq.~(\ref{boson}).
Since the Hamiltonian contains only the spin-conserving products
\begin{equation} \label{a1}
A_{\pm\pm} = \eta_{r\alpha\sigma} 
\eta_{\pm r\pm \alpha\sigma}\;,
\end{equation}
these can be represented using standard Pauli matrices \cite{egger97,nfoot2}.
Besides $A_{++}= 1$, we choose
\begin{equation} \label{a2}
A_{+-}=i\alpha\sigma_x, \;A_{-+}=ir\alpha\sigma_z, \;
A_{--}=-i r\sigma_y\;.
\end{equation}
To show that Eq.~(\ref{a2}) leads to a valid
representation for the Klein factors,
we have to check all possible products
of $A_{\pm\pm}$ with each other. For instance, the
product 
\begin{equation} \label{exm}
A_{+-}A_{-+} = \eta_{r\alpha\sigma}\eta_{r-\alpha\sigma}
\eta_{r\alpha\sigma}\eta_{-r\alpha\sigma}
\end{equation}
gives
$(i\alpha \sigma_x) (ir\alpha \sigma_z ) = i r \sigma_y$
 according to Eq.~(\ref{a2}).
On the other hand, using the anticommutator relation
(\ref{majalg}) gives for Eq.~(\ref{exm})  
$-\eta_{r-\alpha\sigma}\eta_{-r\alpha\sigma}$, which
is again $ir\sigma_y$ from Eq.~(\ref{a2}).
Similarly, all other products are consistent with the
algebra (\ref{majalg}), and one can indeed use the
representation (\ref{a1}) with Eq.~(\ref{a2}).
Of course, our choice (\ref{a2}) is not unique, 
and one can find other possibilities. However, 
final expressions for physically observable quantities
do not depend on this choice.

The bosonized expressions for the various terms
in Eq.~(\ref{ham1})  read \cite{foot}
\begin{eqnarray} \label{bh0}
 H_0 &=& \sum_{j\delta} \frac{v_{j\delta}}{2} 
\int dx \left[ K_{j\delta} ( \partial_x \phi_{j\delta})^2
+ K^{-1}_{j\delta} (\partial_x \theta_{j\delta})^2 \right]\\
\label{bfs0}
H_{\alpha{\rm FS}}^{(0)} &=& \frac{2}{\pi} \int dx dx' \;
\partial_x\theta_{c+}(x) V_0(x-x') \partial_{x'} \theta_{c+}(x')
 \\ \nonumber
 H_{\alpha{\rm FS}}^{(1)} &=& \frac{f}{(\pi a)^2} 
\int dx \; [ -\cos(\sqrt{4\pi} \, \theta_{c-} ) 
\cos(\sqrt{4\pi} \, \theta_{s-} ) \\ 
\label{bfs1} &-&\cos(\sqrt{4\pi} \, \theta_{c-} ) 
\cos(\sqrt{4\pi} \, \theta_{s+} ) \\
\nonumber &+&\cos(\sqrt{4\pi} \, \theta_{s+} ) 
\cos(\sqrt{4\pi} \, \theta_{s-} )]\\
\nonumber H_{\alpha{\rm BS}} &=& \frac{b}{(\pi a)^2} \int dx \, [
\cos(\sqrt{4\pi} \, \theta_{c-} ) \cos(\sqrt{4\pi} \, 
\theta_{s-} ) \\ \label{bbs} &+&
\cos(\sqrt{4\pi} \, \theta_{c-} ) \cos(\sqrt{4\pi} \, \phi_{s-} )\\
\nonumber &+&\cos(\sqrt{4\pi} \, \theta_{s-} ) 
\cos(\sqrt{4\pi} \, \phi_{s-} )]\;.
\end{eqnarray}
The correct signs for the various terms in Eqs.~(\ref{bfs1})
and (\ref{bbs}) are crucial and necessitate a correct choice
for the Majorana fermion products, see Eq.~(\ref{a2}).
Although bosonization of Eq.~(\ref{h02}) gives $K_{j\delta}=1$ in
Eq.~(\ref{bh0}),
interactions renormalize these parameters.
In particular, in the long-wavelength limit,
 $H_{\alpha{\rm FS}}^{(0)}$ 
can be incorporated into $H_0$ by putting 
\begin{equation}\label{Kdef}
K_{c+}= K = \left \{1+ 4\widetilde{V}_0(k\simeq 0)/\pi v \right\}^{-1/2}
 \leq 1 \;,
\end{equation}
while for all other channels [$(j\delta) \neq (c+)$], the coupling
constant $f$  gives rise to the renormalization
\begin{equation}\label{rest}
K_{j\delta}= 1+  f /\pi v \geq 1  \;.
\end{equation}
The corresponding renormalization
of $K_{c+}$ due to $f\neq 0$ can be neglected against the large 
effect of Eq.~(\ref{bfs0}).
The velocities of the different modes in Eq.~(\ref{bh0})
are then given  by
\begin{equation}\label{vv}
v_{j\delta}=v / K_{j\delta} \;.
\end{equation}
Clearly, the charged $(c+)$ mode propagates with significantly
higher velocity than the three neutral modes. 
There is a further renormalization of the values
given by Eq.~(\ref{vv}) due to the $\alpha$BS contribution. 
However, this effect is very small and does not affect the
power laws discussed below. We shall therefore neglect 
velocity renormalizations beyond Eq.~(\ref{vv}) in the
following.

For the long-ranged interaction (\ref{un2}), 
the logarithmic singularity in Eq.~(\ref{Kdef})
has a natural infrared cutoff at $k=2\pi/L$ due to the finite
length of the nanotube. For temperatures $k_B T\gg \hbar v/L$, 
thermal effects provide a higher cut-off. 
Considering  low temperatures, $k_B T\ll\hbar v/L$,
we then obtain
\begin{equation} \label{longr}
K = \left\{ 1+\frac{8e^2}{\pi\kappa \hbar v} [\ln(L/2\pi R) + 0.51 ]
\right\}^{-1/2} \;,
\end{equation}
which predicts $K\to 0$ for $L\to \infty$. Since $\hbar c/e^2\simeq 137$,
we  get with $v=8\times 10^5$ m/sec the estimate
$e^2/\hbar v = (e^2/\hbar c) (c/v) \approx 2.7$, and
therefore $K\simeq 0.18$ for the $L=3\mu$m tube of Ref.\cite{tans}. 
Quite generally, the parameter $K$ gives
the appropriate measure of the correlation strength
\cite{voit,book}.
The rather small value of $K$ found here
implies that a single-wall armchair nanotube is
a very strongly correlated system, which should display pronounced
non-Fermi liquid effects.

\section{Renormalization group analysis}
\label{sec:rg}

The Hamiltonian given by Eqs.~(\ref{bh0})--(\ref{bbs}) 
does not allow for an exact solution.  To proceed,
we investigate the nonlinear terms 
associated with the coupling constants $f$ and $b$ by using
the perturbative RG method \cite{cardy}. 
Since there is no renormalization in the charged $(c+)$
sector, in the following $(j\delta)$ 
includes only the combinations $(c-), (s+)$, and $(s-)$.
There are eight scaling operators 
perturbing the critical $b=f=0$ model
that need to be taken into account. They are given by
\begin{eqnarray}
V_1 &=& \frac{1}{\pi} \cos(\sqrt{4\pi}\theta_{c-}) 
\cos(\sqrt{4\pi}\theta_{s-})\;, \\
V_2 &=& \frac{1}{\pi} \cos(\sqrt{4\pi}\theta_{c-}) 
\cos(\sqrt{4\pi}\theta_{s+})\;, \\
V_3 &=& \frac{1}{\pi} \cos(\sqrt{4\pi}\theta_{s-}) 
\cos(\sqrt{4\pi}\theta_{s+})\;, \\
V_4 &=& \frac{1}{\pi} \cos(\sqrt{4\pi}\theta_{c-}) 
\cos(\sqrt{4\pi}\phi_{s-})\;, \\
V_5 &=& \frac{1}{\pi} \cos(\sqrt{4\pi}\theta_{s+}) 
\cos(\sqrt{4\pi} \phi_{s-}) \;, \\
V_{j\delta} &=& \frac12 \left[ - (\partial_x \phi_{j\delta})^2
+ (\partial_x\theta_{j\delta})^2 \right] \;,
\end{eqnarray}
with associated scaling fields (coupling constants) $g_i(\ell)$ (where
$i=1,\ldots,5$) and $g_{j\delta}(\ell)$. Here $d\ell=-d\ln \omega_c$
denotes the standard RG flow parameter, i.e., one decreases the high-energy
bandwidth cutoff $\omega_c$ and compensates this decrease by
adjusting the coupling constants.  
Within the reach of the perturbative RG,
the coupling constants $g_{j\delta}$ are related
to the $K_{j\delta}$ parameters by $K_{j\delta}=1-2g_{j\delta}$.
The initial values of the coupling constants are 
\begin{eqnarray*}
g_1(0) &=& (b-f)/\pi v \;, \\
g_2(0) &=& -g_3(0)= 2g_{j\delta}(0) =-f/\pi v \;, \\
g_4(0) &=& b/\pi v \;, \\
g_5(0) &=& 0 \;.
\end{eqnarray*}
The operator $V_5$ is not present in the original
Hamiltonian
but will be generated during the renormalization process.
Furthermore, we have omitted the operator 
\[
V' = \frac{1}{\pi} \cos(\sqrt{4\pi}\theta_{s-}) 
\cos(\sqrt{4\pi}\phi_{s-})\;,
\]
which is present in Eq.~(\ref{bbs}). This operator stays
exactly marginal in all orders and thus decouples completely
from the problem.  In fact, by using the Majorana
refermionization procedure in Sec.~\ref{sec:maj}, this
term is seen to vanish.

It is  straightforward to derive the second-order
RG equations, e.g., by using poor man's scaling  
or, more elegantly, the operator product 
expansion  technique \cite{cardy}. 
The resulting equations read
\begin{eqnarray}\label{rgfirst}
\frac{dg_1}{d\ell} &=& -g_2 g_3 + g_1 ( g_{c-}+g_{s-} )\;, \\
\frac{dg_2}{d\ell} &=& -g_1 g_3 -g_4 g_5 + g_2 ( g_{c-}+g_{s+} )\;, \\
\frac{dg_3}{d\ell} &=& -g_1 g_2 + g_3 ( g_{s+}+g_{s-} )\;, \\ \label{g4}
\frac{dg_4}{d\ell} &=& -g_2 g_5 + g_4 ( g_{c-}-g_{s-} )\;, \\ \label{g5}
\frac{dg_5}{d\ell} &=& -g_2 g_4 + g_5 ( g_{s+}-g_{s-} )\;, \\
\frac{dg_{c-}}{d\ell} &=& \frac12 ( g_1^2 + g_2^2 + g_4^2)\;, \\
\frac{dg_{s+}}{d\ell} &=& \frac12 ( g_2^2 + g_3^2 + g_5^2)\;, \\ 
\label{rglast}
\frac{dg_{s-}}{d\ell} &=& \frac12 ( g_1^2 + g_3^2 - g_4^2 - g_5^2)\;.
\end{eqnarray}
We shall first study these equations in two limiting cases,
namely either $b=0$ or $f=0$.

Let us start with the case $b=0$, where the coupling constants
$g_4$ and $g_5$ stay identically zero. Putting
\[
G = -g_1= -g_2 = g_3 
\]
with the initial
condition $G(0)=f/\pi v >0$, and 
\[
\bar{G} = 2 g_{c-} = 2 g_{s-} = 2 g_{s+} 
\]
with $\bar{G}(0)=-f/\pi v <0$, the  RG equations read
\[
\frac{dG}{d\ell} =  - G^2 + \bar{G} G \;,\quad
\frac{d\bar{G}}{d\ell}=   2 G^2  \;.
\]
This permits the solution $G=-\bar{G}$ obeying 
$dG/d\ell =  - 2 G^2$. Since $G(0)>0$, all coupling
 constants flow to zero as $\ell\to \infty$,
\[
G(\ell) = \frac{G(0)}{1+2G(0) \ell} \;.
\]
Therefore, for $b=0$, the contribution $H_{\alpha{\rm FS}}^{(1)}$
in Eq.~(\ref{bfs1}) is {\em marginally irrelevant}. 

Next we turn to the case $f=0$, where one can immediately
put $g_2=g_3=g_5=g_{s+}= 0$.  Furthermore, we are free to
choose the isotropic solution
\[
g = g_1 = g_4 \;,
\]
with the initial condition $g(0)=b/\pi v>0$. Then 
one can also put $g_{s-}=0$ and is left with 
the Kosterlitz-Thouless equations
\begin{equation} \label{kosterlitz}
\frac{dg}{d\ell}= g g_{c-} \;, \quad
\frac{dg_{c-}}{d\ell} =   g^2  \;.
\end{equation}
Therefore both $g$ and $g_{c-}$  will flow to strong coupling, and 
the backscattering part $H_{\alpha{\rm BS}}$  is {\em
marginally relevant}.
The physical picture emerging for the special case $f=0$
is elucidated in Sec.~\ref{sec:maj}.
From this analysis and the numerical solution of the
general RG equations (\ref{rgfirst})--(\ref{rglast}) for $f\neq 0$, 
it is apparent that $b$ is 
always a marginally relevant coupling constant.

The next question to be addressed is whether 
the initially irrelevant $H_{\alpha{\rm FS}}^{(1)}$ becomes
relevant near the emerging strong coupling fixed point.
To study this issue, let us assume $b\gg f$ and linearize the
RG equations in the initially small coupling constants  $g_2, 
g_3, g_5$, and $g_{j\delta}$.  Then $g_{s+}$ decouples,
and we introduce the new coupling constants
\[
g=(g_1+g_4)/2 \;,\quad \bar{g} = g_1-g_4 \;,
\]
with  initial values $g(0)=b/\pi v$ and $\bar{g}(0)=-f/\pi v$.
Thereby we recover the Kosterlitz-Thouless equations (\ref{kosterlitz}),
so that the above strong-coupling behavior of $b$ is basically
unaltered, i.e., the coupling constants $g$ and $g_{c-}$
flow to strong coupling,
$g\to g^*>0$ and $K_{c-}\to K_{c-}^*<1$. The remaining 
RG equations  read
\begin{eqnarray*}
\frac{d\bar{g}}{d\ell} &=&2 g g_{s-}\;,\\
\frac{dg_2}{d\ell} &=& -g (g_3+g_5)\;,\\
\frac{dg_3}{d\ell} &=& \frac{dg_5}{d\ell}= - g g_2\;,\\
\frac{dg_{s-}}{d\ell} &=& \bar{g} g\;.
\end{eqnarray*}
Therefore, $\bar{g}(\ell)$ also reaches the strong coupling regime,
$\bar{g}\to \bar{g}^*<0$, and similarly, $K_{s-}\to
K_{s-}^*>1$.  Furthermore, it
follows readily that $g_2, g_3$ and $g_5$ flow
to strong coupling (with $g_3-g_5$ staying constant).
We conclude that terms associated with the
coupling constant $f$ will become relevant near the strong-coupling
point $b\to  b^*$, which therefore represents only an intermediate
fixed point.  

\begin{figure}
\epsfysize=10cm
\epsffile{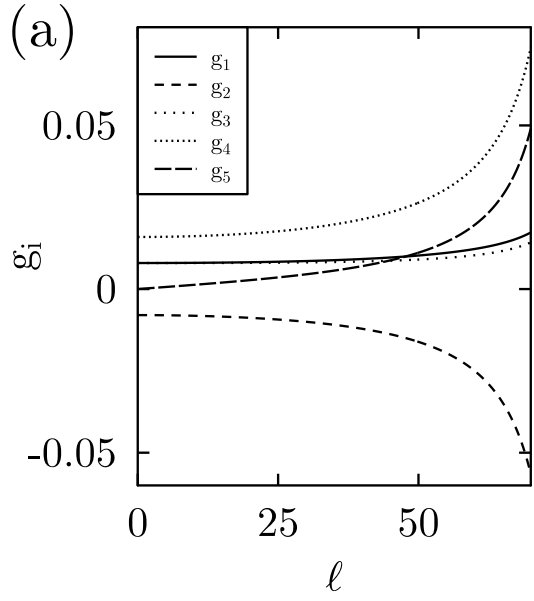}
\epsfysize=10cm
\epsffile{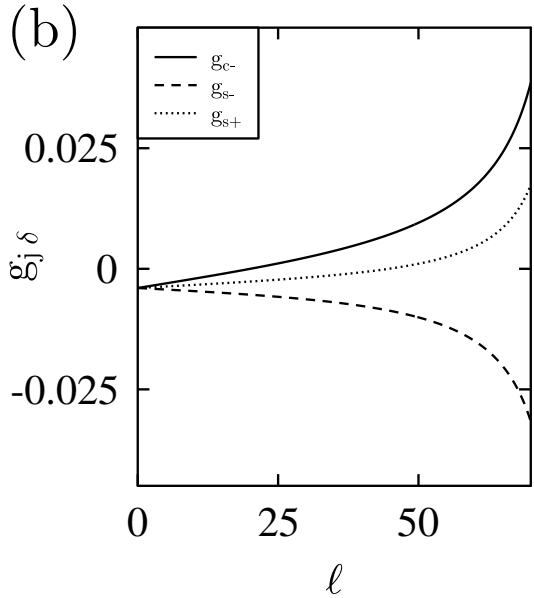}
\caption[]{\label{fig4}  RG flow of the coupling 
constants for $f=0.025$ and $b=0.05$.
Units are such that $a=1$. Following Eqs.~(\ref{gam}) and
(\ref{bbb}), these initial values apply to a $(10,10)$ SWNT
with externally unscreened Coulomb interactions.
}
\end{figure}

\begin{figure}
\epsfysize=10cm
\epsffile{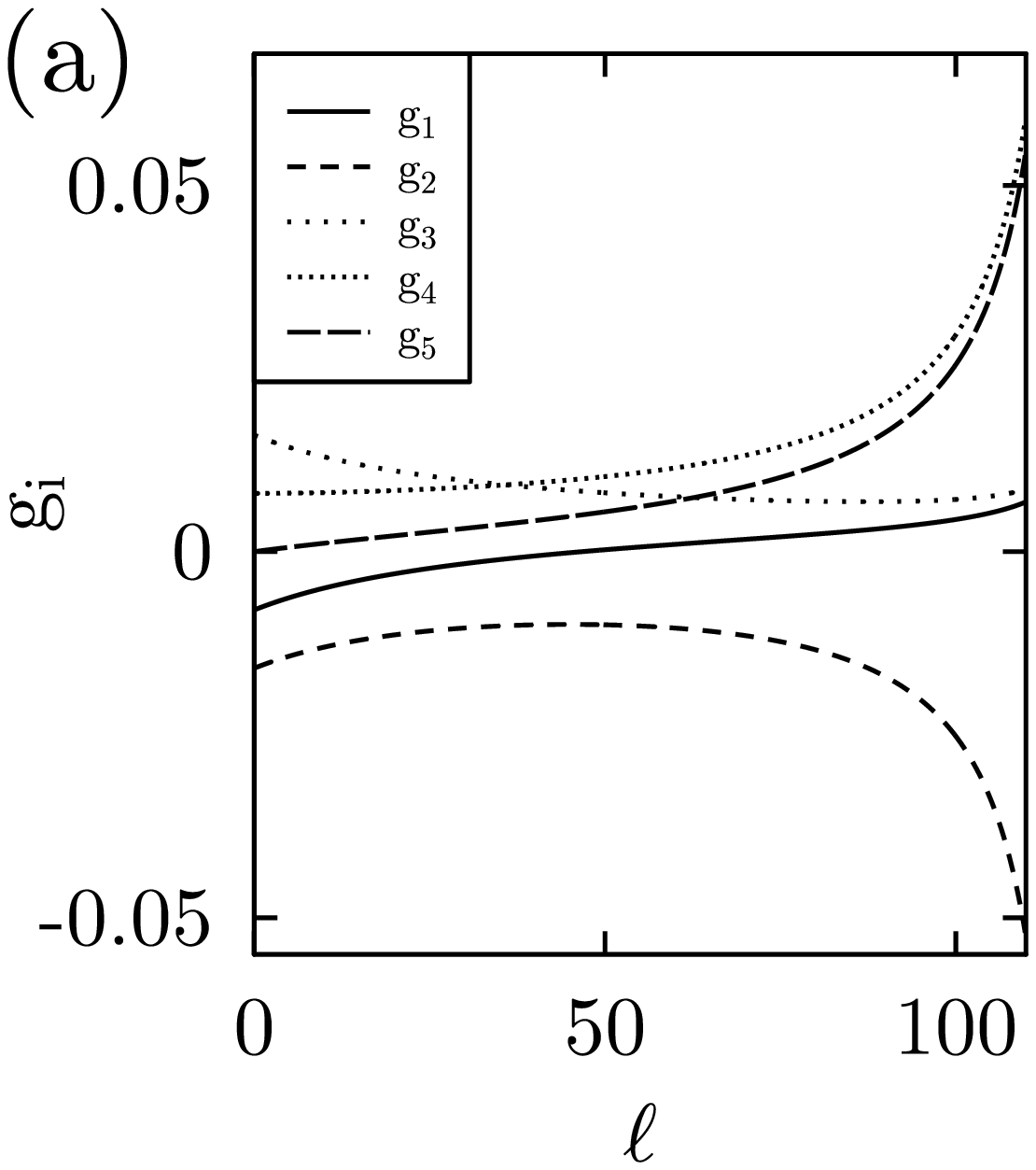}
\epsfysize=10cm
\epsffile{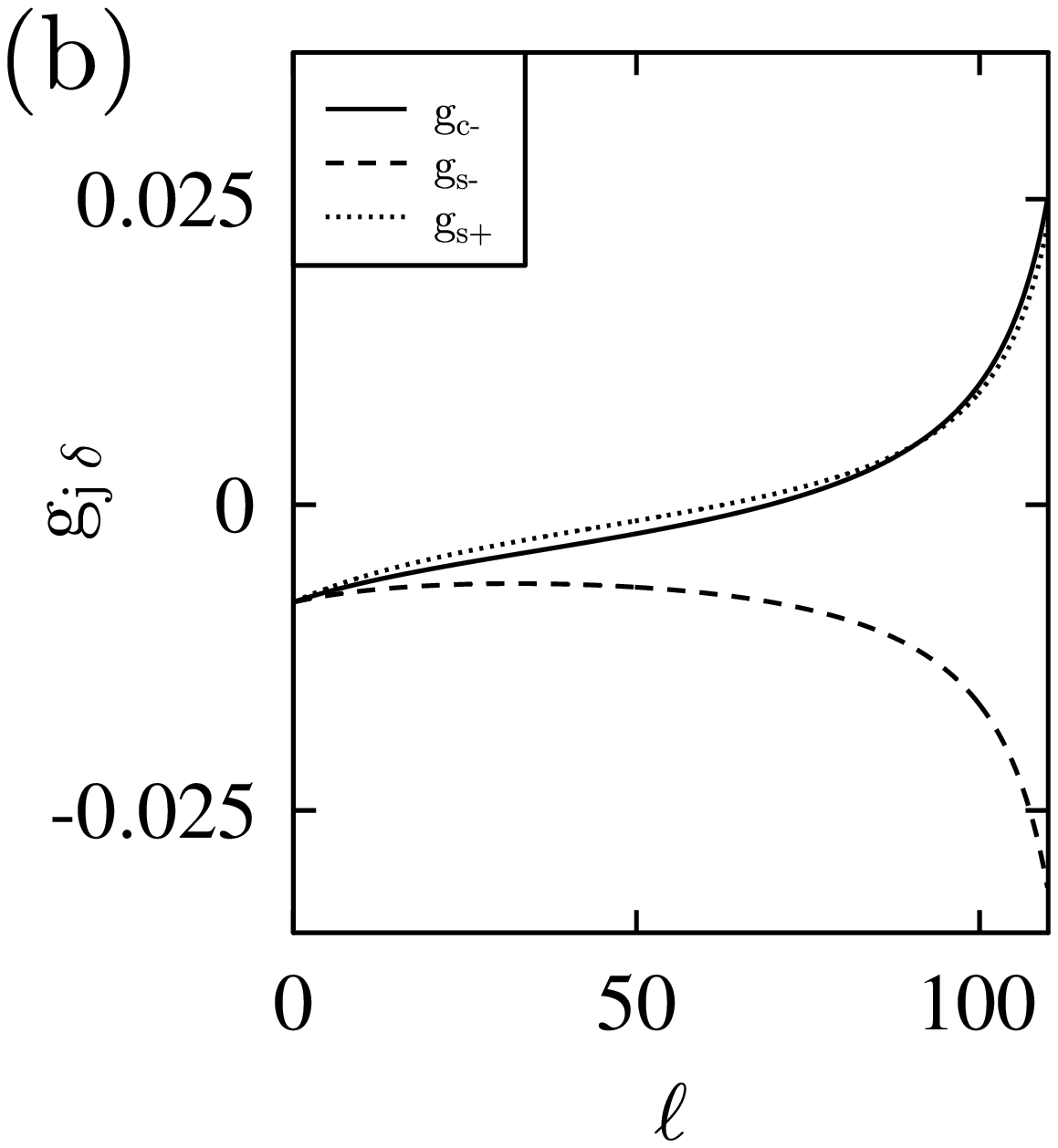}
\caption[]{\label{fig5} 
Same as Fig.~\ref{fig4}, for $f=0.05$ and $b=0.025$.
}
\end{figure}

This conclusion remains valid if the assumption $b\gg f$ is relaxed.
In Fig.~\ref{fig4}, the numerical integration of the RG flow
equations (\ref{rgfirst})--(\ref{rglast}) is shown for 
realistic values of the coupling constants $b$ and $f$.
Clearly, all coupling constants go to  the strong-coupling
regime, albeit $g_4$ associated with the 
backscattering contribution is dominant. 
Only at sufficiently low energy scales, i.e., for large values of 
the flow parameter $\ell$, the
nonlinear forward scattering term $H_{\alpha{\rm FS}}^{(1)}$
becomes important.  Finally, in Fig.~\ref{fig5} we show 
the RG flow for different initial conditions corresponding to
$f>b$.  The coupling constants $g_2$ and $g_3$ associated with 
the forward scattering contribution are initially irrelevant.
However, they eventually become relevant and flow to strong coupling
as $\ell \to \infty$.       
In the end, all coupling constants flow into the strong-coupling
regime again.

\section{Majorana refermionization}
\label{sec:maj}

We start this section by presenting a solution for the special
case $f=0$.  This solution proceeds by Majorana refermionization
and is similar to the one discussed by Schulz
in the context of the standard two-chain problem \cite{schulz}.
For $f=0$, the nonlinearity affects only the relative ($\delta=-$)
channels, for which the bosonized Hamiltonian reads from 
Eqs.~(\ref{bh0}) and (\ref{bbs})
\begin{eqnarray} \label{hamnew}
H(c-,s-) &=&    \frac{v}{2} \int dx \sum_{j=c,s} \left[
 ( \partial_x \phi_{j-})^2
+  (\partial_x \theta_{j-})^2 \right]\\  \nonumber
  &+& \frac{b}{(\pi a)^2} \int dx \, [
\cos(\sqrt{4\pi} \, \theta_{c-} ) \cos(\sqrt{4\pi} \, 
\theta_{s-} ) \\ \nonumber
 &+&
\cos(\sqrt{4\pi} \, \theta_{c-} ) \cos(\sqrt{4\pi} 
\, \phi_{s-} )\\  \nonumber
&+&\cos(\sqrt{4\pi} \, \theta_{s-} ) 
\cos(\sqrt{4\pi} \, \phi_{s-} )]\;.
\end{eqnarray}
Let us define new effective 
fermion operators  for the relative 
charge and spin $(j=c,s)$ channels. Their right- and left-moving
components $(p=\pm=R,L)$ can be written in terms of the bosonic
phase fields,
\[
\psi_{jp} = \frac{\eta_{jp}}{\sqrt{2\pi a}} \exp\{
-i\sqrt{\pi} ( p\theta_{j-} + \phi_{j-} )\} \;.
\]
Then we have
\begin{eqnarray} \label{cosform}
\cos[\sqrt{4\pi} \theta_{j-}] &=& -  \eta^{}_{jR} \eta^{}_{jL} \,
\pi a \left( \psi^\dagger_{jR}
\psi^{}_{jL} - \psi^\dagger_{jL} \psi^{}_{jR} \right) \;, \\ \nonumber
\cos[\sqrt{4\pi} \phi_{j-}] &=& -\eta^{}_{jR}\eta^{}_{jL} \,
\pi a \left( \psi^\dagger_{jR}
\psi^\dagger_{jL} - \psi^{}_{jL} \psi^{}_{jR} \right) \;.
\end{eqnarray}
Next we express these two (complex) Dirac
 fermions in terms of 
four (real) Majorana fermions $\xi_{jp}(x)$.
Here $j=1,2$ corresponds to the $(s-)$ channel,
and $j=3,4$ to $(c-)$,
\begin{eqnarray*}
\psi_{s,R/L} &=& \frac{1}{\sqrt{2}} \left( \xi_{1,R/L}(x) +
i \xi_{2,R/L}(x) \right) \;, \\
\psi_{c,R/L} &=& \frac{1}{\sqrt{2}} \left( \xi_{3,R/L}(x) +
i \xi_{4,R/L}(x) \right) \;.
\end{eqnarray*}
The Majorana fermion operators obey the algebra
\begin{equation}
[ \xi_{jp}(x), \xi_{j'p'}(x') ]_+ = \delta_{jj'} \delta_{pp'}
\delta(x-x') \;,
\end{equation}
and by  using  Eq.~(\ref{cosform}) we find 
\begin{eqnarray} \nonumber
\cos[\sqrt{4\pi} \theta_{s-}] &=& -i\pi a \left( \xi_{1R}
\xi_{1L} + \xi_{2R} \xi_{2L} \right ) \;, \\  \label{cosf2}
\cos[\sqrt{4\pi} \phi_{s-}] &=& -i\pi a \left( \xi_{1R}
\xi_{1L} - \xi_{2R} \xi_{2L} \right )\;, \\ \nonumber
\cos[\sqrt{4\pi} \theta_{c-}] &=& i\pi a \left( \xi_{3R}
\xi_{3L} + \xi_{4R} \xi_{4L} \right ) \;,
\end{eqnarray}
where the Klein factors have been chosen as
$\eta^{}_{cR}\eta^{}_{cL}=-\eta^{}_{sR}\eta^{}_{sL}=-i$.
This choice is dictated by the condition that the
interaction term  is marginally relevant, see below.

As already mentioned,  one can see from this representation that
the operator
\[
\cos[\sqrt{4\pi}\theta_{s-}] \cos [\sqrt{4\pi} \phi_{s-}]
\]
can effectively be put to zero. To be precise, 
upon point splitting \cite{book,cardy}, it
yields highly irrelevant operators involving spatial
derivatives of the Majorana fields.
The corresponding term in Eq.~(\ref{hamnew}) can
thus be omitted.  Refermionization then yields 
[we put $v=1$ in the intermediate steps]
\begin{eqnarray} \nonumber
H(c-,s-) &=& - \frac{i}{2} \sum_{j=1}^4 \int dx\,
 \left( \xi_{jR}\partial_x
\xi_{jR} - \xi_{jL} \partial_x \xi_{jL} \right)  \\
&+& 2 b \int dx\, \left(\xi_{3R}
\xi_{3L} + \xi_{4R}\xi_{4L}\right) \xi_{1R}\xi_{1L}  \;.
\label{maj}
\end{eqnarray}

In order to better understand the above model,
it is instructive to rewrite it in terms
of the following current operators,
\begin{eqnarray}
I^x_R(x)&=&i\xi_{1R}(x)\xi_{3R}(x) 
\;,\nonumber\\
I^y_R(x)&=&-i\xi_{1R}(x)\xi_{4R}(x) \;,\label{currents}\\
I^z_R(x)&=&i\xi_{3R}(x)\xi_{4R}(x)\;,\nonumber
\end{eqnarray}
and analogously for the left-movers.
These currents obey the $SU(2)$ level-2 Kac-Moody algebra \cite{book},
\begin{equation} \label{Kac-Moody}
[I^a(x),I^b(y)]_-=i\epsilon^{abc}\delta(x-y)I^c(x)+
\frac{i\delta_{ab}}{2\pi}\delta'(x-y) \;.
\end{equation}
In terms of the operators (\ref{currents}),
the interacting part of the Hamiltonian (\ref{maj})
takes the form
\[
\int dx \left\{ g_\perp\left[I^x_R(x)I^x_L(x)+
I^y_R(x)I^y_L(x)\right]+g_zI^z_R(x)I^z_L(x)\right\}\;.
\]
The transverse coupling is $g_\perp=2b$, while
the longitudinal coupling $g_z$ is initially absent
but will be generated under the renormalization.
We thus can identify the model (\ref{maj}) as
the $SU_2(2)$ anisotropic Wess-Zumino-Witten
model, see, e.g., Ref.\cite{book}.
From the structure of the commutation relations,
it immediately follows that the coupling constants
$(g_\perp, g_z)$ obey the Kosterlitz-Thouless equations
\begin{equation}
\frac{dg_\perp}{d\ell}=\frac{1}{2\pi}g_\perp g_z\;,
\;\;\;
\frac{dg_z}{d\ell}=\frac{1}{2\pi}g_\perp^2\;.
\label{currentsRG}
\end{equation}
Using the operator identity
\begin{eqnarray}
\left \{ \frac{1}{\pi a}\cos[\sqrt{4\pi}\theta(x)]\right\}^2 &=&
\frac{1}{(\pi a)^2}\cos[\sqrt{16\pi}\theta(x)]\nonumber\\
&-&\frac{1}{\pi}(\partial_x\theta)^2+{\rm const.}\;,
\label{identity}
\end{eqnarray}
we conclude that the longitudinal coupling corresponds
to a  renormalization of $K_{c-}$,
with $g_z=2\pi g_{c-}$.
Using this identification we recover the flow equations
(\ref{kosterlitz}) and highlight the underlying reason
for their Kosterlitz-Thouless structure.

Of course, for $b>0$, the model flows to strong
coupling, rendering the Majorana fields $\xi_1$,
$\xi_3$ and $\xi_4$ massive. Notice, however, that
the Majorana fermion $\xi_2$ stays {\em massless},
so that the $(s-)$ sector carries one massive ($j=1$) and
one massless branch. This behavior is due to the symmetric 
appearance of the dual fields $\theta_{s-}$ and $\phi_{s-}$ 
(self-duality) in
Eq.~(\ref{hamnew}). By virtue of the Heisenberg uncertainty
relation, it is impossible to completely pin a self-dual field.
However, the $(c-)$ sector is fully massive.

The masses of the massive branches can now safely be calculated
by Majorana mean-field theory.  We mention in passing that
a standard self-consistent harmonic treatment \cite{book}
 does not apply
in this case since we have to deal with a {\em marginally} relevant
perturbation.   In the mean-field approximation,
 the interaction term in Eq.~(\ref{maj}) is written as
\begin{equation} \label{mf}
-2i b \int dx \, \left( c_1 \left [ \xi_{3R} \xi_{3L} + 
\xi_{4R}\xi_{4L} \right] + 2c_3 \,\xi_{1R}\xi_{1L} \right) \;,
\end{equation}
with the ground-state averages
\begin{eqnarray*}
 c_1 &=& i \langle \xi_{1R} \xi_{1L} \rangle \;,\\
c_3 &=& c_4 = i\langle \xi_{3R} \xi_{3L} \rangle  =
i\langle \xi_{4R} \xi_{4L} \rangle \;.
\end{eqnarray*}
At this point it is straightforward to 
diagonalize the mean-field Hamiltonian.
This yields the self-consistency relations
\begin{eqnarray*}
c_1 &=& \frac{2bc_3}{\pi} \ln ( \omega_c/4 b c_3) \;, \\
c_3 &=& \frac{bc_1}{\pi} \ln ( \omega_c/2 b c_1) \;,
\end{eqnarray*}
where $\omega_c = 7.4$ eV is the bandwidth of the $\pi$ 
electrons \cite{hamada}.
The solution to these relations is $c_3\simeq c_1/\sqrt{2}$
with 
\[
c_1 \simeq \frac{\omega_c}{2b} \exp[-\pi / \sqrt{2} b]\;.
\]
Comparing Eq.~(\ref{mf}) with the usual mass term 
$-i m \int dx \xi_R \xi_L$ yields the masses $m_j$ of the
massive Majorana fermion branches $j=1,3,4$.  Apart from a 
factor $\sqrt{2}$, these masses are all equal and given
by 
\begin{equation}\label{massb}
m_b = \omega_c \exp[-\pi v / \sqrt{2} b] \;.
\end{equation}
Since the coupling constant $b$ appears in the 
exponent of an exponentially small quantity,
it is quite difficult to estimate the value of $m_b$. 
Taking the value (\ref{bbb}) for $b$ yields an order-of-magnitude
estimate for the associated temperature scale, $T_b \approx
0.1$ mK.  For well-screened interactions, however, $T_b$ 
can be orders of magnitude larger.

To proceed, since the $(c-)$ sector is fully massive,
we now effectively put $\theta_{c-}=0$ in Eq.~(\ref{hamnew}).
Other degenerate solutions give the same physical results.
Then we are confronted with the interaction Hamiltonian
\[
\frac{m_b}{2\pi a} \int dx \left( \cos[\sqrt{4\pi} \theta_{s-}]
+ \cos[\sqrt{4\pi} \phi_{s-} ] \right) \;.
\]
Self-duality implies that 
 $\sin[\sqrt{\pi} \theta_{s-}]$ and 
 $\sin[\sqrt{\pi} \phi_{s-}]$ must have the same scaling dimension.
Similarly, the scaling dimensions of $\cos[\sqrt{\pi}\theta_{s-}]$ 
and $\cos[\sqrt{\pi} \phi_{s-}]$ must coincide. 
 Although the fields $\theta_{s-}$ and
$\phi_{s-}$ are not pinned, they still have a tendency
to be pinned. Since $m_b>0$, this implies that 
the values of $\sin[\sqrt{\pi} \theta_{s-}]$ and 
$\sin[\sqrt{\pi} \phi_{s-}]$ are close to 1, while
the corresponding $\cos$-operators approach zero.
Therefore the $\sin$-operators will be correlated stronger
than the $\cos$-operators.

These simple arguments can be made quantitative as follows.
The $(s-)$ sector is described by two decoupled
Hamiltonians, $H=H_1[\xi_1]+H_2[\xi_2]$, for the Majorana
fermions $\xi_1$ and $\xi_2$, 
\begin{eqnarray}\label{maj2} 
H_j[\xi_j] &=& -\frac{i}{2} \int dx \left ( 
\xi_{jR}\partial_x \xi_{jR} - 
\xi_{jL}\partial_x \xi_{jL} \right) \\ \nonumber
&-& i m_j \int dx \, \xi_{jR} \xi_{jL} \;,
\end{eqnarray}
where the masses are given by $m_1=m_b$ and $m_2=0$.
We now exploit the well-known correspondence between
the 2D classical Ising model and
1D Majorana fermions \cite{book,itzykson}.
The fermion mass $m_j$ sets the relevant energy scale, 
where the correlation length of the related Ising model 
should be proportional to $m_j$. Therefore the 
Ising model related to $\xi_2$ will be critical ($T=T_c$),
while the Ising model for $\xi_1$ is above criticality (since $m_1>0$,
we have $T>T_c$).
Denoting the standard order operators for the two Ising models as 
$\sigma_{1,2}$, and disorder operators as $\mu_{1,2}$,
the {\em fusion rules} \cite{book,itzykson}
yield the correspondence
\begin{eqnarray} \label{fusion}
\cos[\sqrt{\pi} \theta_{s-}] &=& \sigma_1 \mu_2 \;,\\ \nonumber
\cos[\sqrt{\pi} \phi_{s-}] &=& \sigma_1 \sigma_2 \;,\\ \nonumber
\sin[\sqrt{\pi} \theta_{s-}] &=& \mu_1 \sigma_2 \;,\\ \nonumber
\sin[\sqrt{\pi} \phi_{s-}] &=& \mu_1 \mu_2\;.
\end{eqnarray}
These fusion rules state that for two given Ising models,  products
of the order/disorder operators determine
the Majorana field operators,
which in turn are composed out of the exponentials 
$\exp[\pm i\sqrt{\pi} \theta_{s-}]$ and
$\exp[\pm i\sqrt{\pi} \phi_{s-}]$.  The relation
(\ref{fusion}) is a valid representation of the
fusion rules, see Ref.\cite{book}.
 
Since the first Ising model is above criticality, the
average of the order operator is zero, $\langle\sigma_1\rangle=0$.
However, the disorder operator $\mu_1$ has a finite average value,
and correlation functions of $\cos[\sqrt{\pi} \theta_{s-}]$
and $\cos[\sqrt{\pi} \phi_{s-}]$ decay exponentially.
From the exact solution of the 2D Ising model \cite{itzykson}
one then immediately obtains
\begin{equation} \label{ising}
\langle \sin[\sqrt{\pi} \theta_{s-}(x)] 
\sin[\sqrt{\pi} \theta_{s-}(x')] \rangle \sim |x-x'|^{-1/4} \;,
\end{equation}
with the same result for the $\sin[\sqrt{\pi} \phi_{s-}]$
correlator. 
The $\sin$-operators therefore contain the Ising operator $\mu_1$
from the off-critical model which has a finite average.
The other Ising operator then comes from a critical
model ($m_2=0$) and leads to the scaling dimension
$\eta=1/8$, see Eq.~(\ref{ising}).  In contrast, the
$\cos$-operators contain the Ising operator $\sigma_1$ 
 with zero expectation value.
Note that for $b=0$ all these operators have scaling
dimension $\eta=1/4$.  The ``halved'' scaling dimension $\eta=1/8$
found for $b>0$ can be traced to the fact that one Majorana
fermion becomes massive while the other stays massless.
In the $(c-)$ sector, correlations of 
$\cos[\sqrt{\pi} \theta_{c-}]$ show long-range order while all
other operators lead to exponential decay.

The results obtained so far by Majorana refermionization
apply only to the intermediate fixed point characterized
by $f=0$.   As discussed
in Sec.~\ref{sec:rg},
while $H_{\alpha{\rm FS}}^{(1)}$ is irrelevant around the
non-interacting fixed point,
it becomes relevant near this intermediate strong-coupling point.
The term $\sim \cos(\sqrt{4\pi} \theta_{s+})
\cos(\sqrt{4\pi} \theta_{s-})$ in Eq.~(\ref{bfs1}) stays marginal,
but the two other terms become relevant with scaling dimension
$\eta=1$.  This renders the $(s+)$ channel massive.
Regarding the $(s-)$ channel, the most 
important  contribution due to $H^{(1)}_{\alpha{\rm FS}}$
comes from 
\[
H_{\alpha{\rm FS}}^{(1)} \simeq - \frac{f}{(\pi a)^2}
\int dx \, \cos[\sqrt{4\pi} \theta_{c-}] \cos[\sqrt{4\pi} \theta_{s-}] \;.
\]
Noting that we have chosen $\theta_{c-}=0$, we can again
employ Majorana refermionization.  The only but important effect of 
this contribution then consists of a 
renormalization of the masses $m_{1,2}$,
\[
m_1 \to (m_b-  m_f) \;, \quad m_2 \to m_f \;,
\]
which breaks the self-duality in the $(s-)$ sector
and thus drives the second Ising model off criticality as well.
Majorana mean-field theory yields
\begin{equation}\label{massf}
m_f \simeq (f/b) \, m_b \;,
\end{equation} 
and the associated temperature scale $T_f\approx (f/b)\, T_b$.
Therefore both the Majorana fermion $\xi_2$ and
the $(s+)$ field  acquire the (generally small) mass $m_f$ due to
the forward scattering contribution. 
At the emerging $T=0$ strong-coupling fixed point, we have
 long-range order in the operators 
\begin{equation}\label{pincond}
\cos[\sqrt{\pi} \theta_{s+}] \;, \quad 
\cos[\sqrt{\pi} \theta_{c-}]\;, \quad \sin[\sqrt{\pi} \phi_{s-}] \;,
\end{equation}
with exponential decay in all other operators 
except those of the critical $(c+)$ sector.  
That the first two operators become long-ranged is 
a direct consequence of the pinning condition $\theta_{c-}=0$.
That $\sin[\sqrt{\pi}\phi_{s-}]$
exhibits long-range order can be easily understood
from Eq.~(\ref{fusion}),
since the second Ising model is now also above criticality. Therefore
both $\langle\mu_1\rangle$ and $\langle\mu_2\rangle$ are finite,
but $\langle\sigma_1\rangle=\langle\sigma_2\rangle=0$.

This analysis for a general interaction potential predicts that
the exponents corresponding to
the first (intermediate) strong-coupling point 
should be observable on
 temperature scales $T_f <  T < T_b$, with a
crossover to a regime $T < T_f$ dominated by the
true $T=0$ fixed point.
For long-ranged interactions, we have  $T_f\approx T_b$, and
the  intermediate fixed point 
and the associated crossover phenomenon are not observable.
However, for screened interactions,
$T_b$ can be significantly larger than $T_f$.
Finally, in the regime $T>T_b$, which also includes the 
temperature range experimentally studied in Ref.\cite{tans},
the physical behavior emerging from our analysis is 
best characterized as that of a {\em Luttinger liquid}\, with an 
additional flavor index. 
In this regime it is justified to neglect the nonlinearities
associated with the coupling constants $b$ and $f$.
 Nanotubes therefore constitute
a realization of the Luttinger liquid at sufficiently
high temperatures.

\section{Susceptibilities and correlation functions}
\label{sec:susc}

With the strong-coupling solution of Sec.~\ref{sec:maj}
at hand, we can now
examine temperature-dependent susceptibilities and
other experimentally accessible quantities. 
Due to the Mermin-Wagner theorem, 
 there can be no ordered state in a
1D system \cite{foot1}.
It is therefore customary to characterize
the physical behavior by the most slowly decaying 
correlation function, which in turn indicates
incipient instabilities.  We have investigated
correlators of charge-density wave (CDW),
spin-density wave (SDW), and superconducting (SC) type.
The results reported in this section are summarized in 
Table \ref{table1}.

One has to carefully distinguish the three temperature regimes 
discussed in Sec.~\ref{sec:maj}, since different
instabilities emerge in different temperature ranges.
Furthermore, we also have to distinguish the spatial
oscillation period of the correlations.  It is apparent
from the dispersion relation in Fig.~\ref{fig1} that
the wavelengths 
\begin{equation}\label{wavelength}
\lambda = \pi/k_F ,\quad \pi/q_F, \quad \pi/(k_F \pm q_F ) 
\end{equation}
could occur in a doped armchair SWNT (we put $q_F>0$ for simplicity). 
Without doping, only the standard $\pi/k_F$ wavelength
is found, which
corresponds to an order operator effectively involving two 1D
fermions at different Fermi points ($\alpha=-\alpha'$).
In the doped case, the much longer
wavelength $\pi/q_F$ arises if the 1D fermions
are at the same Fermi point but move in opposite directions 
($r=- r'$).
Finally, the wavelength $\pi/(k_F\pm q_F)$
corresponds to the mixed situation ($\alpha=-\alpha', r=-r'$).  The possible
simultaneous occurrence of different wavelengths is a remarkable 
feature of doped nanotubes which has no analogue in the standard
two-chain problem.  
This phenomenon has its ultimate origin in the unique 
band structure of graphite.

\subsection{CDW correlations}
\label{sec:cdw}

In calculating the CDW correlations, we first need to 
find the bosonized representation of the $y$-integrated
[this only amounts to a factor $2\pi R$] density operator $q(x)$.
It can be derived from Eq.~(\ref{expa}),
\begin{eqnarray} \nonumber
q(x) &=& \int dy \sum_\sigma \Psi_\sigma^\dagger(\bbox{x})
\Psi_\sigma^{} (\bbox{x}) \\ \label{densop}
&=& \sum_{pp'\alpha\alpha'\sigma}  \int dy
\varphi^*_{p\alpha}(\bbox{x}) \varphi^{}_{p'\alpha'}(\bbox{x})
\, \psi^\dagger_{p\alpha\sigma}(x) \psi^{}_{p'\alpha'
\sigma}(x) \;.
\end{eqnarray}
There is first a ``slow'' component $\rho(x)$ due to 
$p=p', \alpha=\alpha'$, whose bosonized form is given
in Eq.~(\ref{tode2}).  Furthermore, there is an
intra-sublattice (CDW0) order parameter contributing in
Eq.~(\ref{densop}),
\[ 
\hat{O}_{CDW0}(x) \sim \sum_{p\alpha\sigma} \psi^\dagger_{p\alpha\sigma} (x)
\psi^{}_{p,-\alpha,\sigma} (x) \; .
\]
Since the Bloch functions for different sublattices are orthogonal,
one might now conclude that $p\neq p'$ cannot give a 
contribution to $q(x)$. However, the corresponding finite
(but generally small) matrix element allowing for a
contribution due to the inter-sublattice (CDW$\pi$) 
order parameter
\[
\hat{O}_{CDW\pi}(x) \sim \sum_{p\alpha \sigma}  
\psi^\dagger_{p\alpha \sigma} (x) \psi^{}_{-p,\pm\alpha,\sigma} (x) 
\]
can be generated by the interactions.  Such a mechanism is well-known
from the study of $4k_F$ components in the density operator
for correlated 1D fermions \cite{schulzlr}.

To make use of the bosonized version, we then
employ the unitary transformation (\ref{rotate}) and find
\begin{eqnarray} \nonumber
\hat{O}_{CDW0} &\sim& 
\sum_{r\alpha\sigma} \widetilde{\psi}^\dagger_{r\alpha\sigma}
\widetilde{\psi}^{}_{r,-\alpha,\sigma} \\ 
\label{ocdw0} &\sim& \sin[\sqrt{\pi}\phi_{c-}+2k_F x]
\cos[\sqrt{\pi} \theta_{c-}]
\\ \nonumber &\times&
\cos[\sqrt{\pi} \theta_{s-}] 
\cos[\sqrt{\pi} \phi_{s-}]  - (\cos \leftrightarrow \sin) \;.
\end{eqnarray}
The correlation function of 
$\hat{O}_{CDW0}$ thus has the wavelength $\lambda=\pi/k_F$
and decays exponentially for $T<T_b$.
In contrast, for $T>T_b$, its scaling dimension is $\eta=1$.
We therefore omit the CDW$0$ mode in the sequel as it exhibits 
a fast subdominant decay for all temperatures.

Turning to the CDW$\pi$ correlations, we find
two contributions,
\begin{eqnarray} \nonumber
 \hat{O}_1 &\sim& \sum_{r\alpha \sigma} (-ir) 
\widetilde{\psi}^\dagger_{r\alpha \sigma}
\widetilde{\psi}^{}_{-r,\alpha,\sigma} \\
\label{ocdw1}
&\sim& \cos[\sqrt{\pi} \theta_{c+}+2q_F x]
\cos[\sqrt{\pi} \theta_{c-}]
\\ \nonumber &\times&
\sin[\sqrt{\pi} \theta_{s+}] 
\sin[\sqrt{\pi} \theta_{s-}]  + (\cos \leftrightarrow \sin) \;,\\
\nonumber 
 \hat{O}_2 &\sim& \sum_{r\alpha \sigma} (-ir) 
\widetilde{\psi}^\dagger_{r\alpha \sigma}
\widetilde{\psi}^{}_{-r,-\alpha,\sigma} 
\\ \label{ocdw2}
&\sim&
\cos[\sqrt{\pi} \theta_{c+}+2q_F x]
\cos[\sqrt{\pi} \phi_{c-}+2k_F x]
\\ \nonumber &\times&
\cos[\sqrt{\pi} \theta_{s+}] 
\cos[\sqrt{\pi} \phi_{s-}]  + (\cos \leftrightarrow \sin) \;.
\end{eqnarray}
The first operator leads to a slowly oscillating correlation function 
with wavelength $\lambda=\pi/q_F$, while the second operator
implies rapid oscillations with $\lambda=\pi/(k_F\pm q_F)$.
Both operators exhibit exponential decay 
at the lowest temperatures,
$T<T_f$. 
In the intermediate temperature regime, $T_f<T<T_b$, 
$\hat{O}_2$ also leads to exponentially decaying correlation functions,
but $\hat{O}_1$ has the scaling
dimension $\eta=(3+2K)/8$.  Finally, in the high-temperature
Luttinger liquid regime, $T>T_b$, the scaling dimension
$\eta = (3+K)/4$ arises for both operators.  For scaling dimension
$\eta$, the equal-time correlation function is
\begin{equation}\label{scalingdim}
\langle \hat{O}_1(x) \hat{O}_1(x')\rangle \sim
\cos[2q_F(x-x')] \, |x-x'|^{-2\eta} \;,
\end{equation}
and the corresponding susceptibility 
has the temperature dependence $\chi(2q_F) \sim T^{2\eta -2}$. 
For correlations of $\hat{O}_2$,  
instead of $2q_F$ one has the wavevector $2(k_F\pm q_F)$.
Since only the slowly oscillating contribution due to 
$\hat{O}_1$  exhibits power-law
behavior for $T_f<T<T_b$,  it is favored
over the $\hat{O}_2$ contribution
by a larger prefactor for $T>T_b$, 
at least for well-screened interactions.

\subsection{Higher-order CDW correlations}
\label{sec:wigner}

There is also a contribution to the density correlation function 
effectively originating from squaring the above order parameters.
Even if the original operators are irrelevant,
the emerging higher-order operators can become relevant.
The analogue of this behavior in the standard two-chain
problem has been interpreted as incipient Wigner
crystal behavior \cite{schulz}, since it is characterized by a wavelength
$\pi/2 k_F$ corresponding to the average 
 electronic spacing.  Such an interpretation is 
not possible in our case, since the characteristic wavevector
is now $4q_F$ or even $8q_F$ instead of the standard value
$4k_F$.

The operator $\hat{O}_{CDW0}^2$ stays always irrelevant,
but the squared CDW$\pi$ operators give important
 contributions. Using Eq.~(\ref{identity}), we find the leading
terms
\begin{eqnarray} \nonumber
\hat{O}_1^2 &\sim& \cos[\sqrt{4\pi}\theta_{c+}+4q_F x] \Bigl\{
-\cos[\sqrt{4\pi} \theta_{s-}]\\  \label{o12} &+&
\cos[\sqrt{4\pi} \theta_{c-}]-\cos[\sqrt{4\pi} \theta_{s+}]
\Bigr\}\;, \\
\nonumber
\hat{O}_2^2 &\sim& \cos[\sqrt{4\pi}\theta_{c+}+4q_F x] \Bigl\{
\cos[\sqrt{4\pi} \theta_{s+}]\\  \label{o22} &+&
\cos[\sqrt{4\pi} \phi_{c-}+4k_F x]+\cos[\sqrt{4\pi} \phi_{s-}]
\Bigr \} \;.
\end{eqnarray}
Here we have kept only  contributions containing the $\theta_{c+}$
field, since these are the only ones which could become relevant.
Since these operators come from products of four 1D fermion
operators,   the Pauli principle allows exactly three
terms for $\hat{O}^2_1$ and $\hat{O}^2_2$, respectively.
The prefactors for the various terms in Eqs.~(\ref{o12}) 
and (\ref{o22}) are non-universal and
depend on the interactions.
The scaling dimension of these operators is $\eta=1+K$ for
$T>T_b$, so that they are always irrelevant in the Luttinger
liquid regime.  Furthermore, for $T<T_b$, we observe 
 that $\hat{O}_2^2$ also produces
only subleading contributions. In effect, we need to consider
only $\hat{O}_1^2$, which indeed becomes a relevant operator 
for $T<T_b$. The scaling dimension
in this temperature regime is $\eta=K$, leading to the
$4q_F$ oscillatory correlations
\begin{equation}\label{cdw2}
\langle \hat{O}^2_{CDW\pi}(x) \hat{O}^2_{CDW\pi}(x')\rangle \sim
\cos[4q_F(x-x')] \, |x-x'|^{-2K} \;.
\end{equation}
This turns out to be the dominant instability at $T<T_b$ 
and $K<1/2$, see Table \ref{table1}.

In the same line of reasoning, there  also 
exists an effective contribution due to  $\hat{O}_{CDW\pi}^4$.
This operator is relevant  for
extremely strong correlations,
\begin{equation}\label{o4}
\hat{O}_{CDW\pi}^4 \sim
 \cos[\sqrt{16\pi}\theta_{c+}+8q_F x] \;,
\end{equation}
with wavelength $\pi/4q_F$
and scaling dimension $\eta=4K$ for all temperatures.
By comparing to the scaling dimension $\eta=(3+K)/4$ of $\hat{O}_1$,
the $\lambda=\pi/4q_F$ CDW$\pi$ state
is seen to represent the dominant incipient instability in the Luttinger
liquid regime $T>T_b$ for very strong but still realistic
 correlations, $K<1/5$.
At temperatures below $T_b$, the $8q_F$ operator (\ref{o4}) leads
to only subdominant contributions and can safely be ignored.

\subsection{SDW correlations}

Next we turn to {\em spin-density wave} correlations.
Due to the underlying $SU(2)$ spin isotropy,
it is sufficient to study the
$s_z$ correlations which can be readily evaluated. 
Similar to the CDW case, we have the order parameters 
\begin{eqnarray*} 
\hat{O}_{SDW0}(x) &\sim& \sum_{p\alpha\sigma} 
\sigma\psi^\dagger_{p\alpha\sigma} (x)
\psi^{}_{p,-\alpha,\sigma} (x) \;, \\
\hat{O}_{SDW\pi}(x) &\sim& \sum_{p\alpha \sigma}  \sigma
\psi^\dagger_{p\alpha \sigma} (x) \psi^{}_{-p,\pm\alpha,\sigma} (x) \;.
\end{eqnarray*}
Again the intra-sublattice SDW0 component
can be ignored as it only leads to $\eta\geq 1$. 
This is apparent from the bosonized 
representation
\begin{eqnarray} \nonumber
\hat{O}_{SDW0} &\sim& 
\sum_{r\alpha\sigma} \sigma \widetilde{\psi}^\dagger_{r\alpha\sigma}
\widetilde{\psi}^{}_{r,-\alpha,\sigma} \\ 
\label{osdw0} &\sim& \cos[\sqrt{\pi}\phi_{c-}+2k_F x]
\cos[\sqrt{\pi} \theta_{c-}]
\\ \nonumber &\times&
\cos[\sqrt{\pi} \theta_{s-}] 
\sin[\sqrt{\pi} \phi_{s-}]  - (\cos \leftrightarrow \sin) \;.
\end{eqnarray}
The important part comes from the SDW$\pi$ order
operator 
\[
\hat{O}_{SDW\pi} \sim \sum_{p\alpha\sigma} \sigma
\psi^\dagger_{p\alpha\sigma} \psi^{}_{-p,\pm\alpha,\sigma}\;.
\]
Rotation to the right/left-moving basis and subsequent
bosonization gives the two operators
\begin{eqnarray} \nonumber
 \hat{O}_a &\sim& \sum_{r\alpha \sigma} (-ir\sigma) 
\widetilde{\psi}^\dagger_{r\alpha \sigma}
\widetilde{\psi}^{}_{-r,\alpha,\sigma} \\ \label{osdw1}
&\sim& \sin[\sqrt{\pi} \theta_{c+}+2q_F x]
\cos[\sqrt{\pi} \theta_{c-}] \\ \nonumber &\times&
\cos[\sqrt{\pi} \theta_{s+}] 
\sin[\sqrt{\pi} \theta_{s-}]  + (\cos \leftrightarrow \sin) \;,\\
\nonumber \hat{O}_b &\sim& \sum_{r\alpha \sigma} (-ir\sigma) 
\widetilde{\psi}^\dagger_{r\alpha \sigma}
\widetilde{\psi}^{}_{-r,-\alpha,\sigma} 
\\ \label{osdw2} &\sim& \cos[\sqrt{\pi} \theta_{c+}+2q_F x]
\sin[\sqrt{\pi} \phi_{c-}+2k_F x] \\ \nonumber &\times&
\cos[\sqrt{\pi} \theta_{s+}] 
\sin[\sqrt{\pi} \phi_{s-}]  + (\cos \leftrightarrow \sin) \;.
\end{eqnarray}
Both operators lead to exponentially decaying correlations
at $T<T_f$. Furthermore, in the regime $T_f<T<T_b$, $\hat{O}_a$
has the same scaling dimension $\eta=(3+2K)/8$ as the CDW$\pi$
operator $\hat{O}_1$ in Eq.~(\ref{ocdw1}), with exponential decay
in $\hat{O}_b$.  Finally,
for $T>T_b$, both $\hat{O}_{a,b}$ have again the same scaling dimension
$\eta=(3+K)/4$ as  $\hat{O}_{1,2}$. 
We then arrive at exactly the same power laws for SDW and CDW
correlations. 

These power laws represent the most slowly decaying correlations
of the system  in the wide temperature range $T>T_f$, 
provided the correlations are not too strong, see Table \ref{table1}.
To discriminate among CDW and SDW correlations,
we have to study prefactors of the power laws. 
From Eqs.~(\ref{ocdw1}) and (\ref{osdw1}), 
since  $\cos[\sqrt{\pi}\theta_{s+}]$
appears in $\hat{O}_a$ instead of 
$\sin[\sqrt{\pi} \theta_{s+}]$ for $\hat{O}_1$,
and because at temperatures $T<T_f$ we have the pinning condition
$\theta_{s+}=0$, see Eq.~(\ref{pincond}),
the SDW amplitude should be larger than the
CDW amplitude also for $T>T_f$. 
The magnetic {\em SDW correlations} then represent the 
{\em dominant instability} of
the nanotube in the temperature  regime $T>T_f$ for moderate 
correlation strength.
  Remarkably,  they are characterized by the
simultaneous presence of the
wavelengths $\lambda=\pi/q_F$ and $\pi/(k_F\pm q_F)$.

Unfortunately, since the intermediate temperature regime is absent 
for externally unscreened interactions, i.e., $T_f\approx T_b$,
it appears to be rather difficult 
to decide which wavelength will eventually be more
important for the experimental setup of Ref.\cite{tans}. 
For a well-screened interaction, based on the above discussion,
we expect that the slow wavelength is more important.
In that case, since $\hat{O}_{SDW\pi}$ can give a
contribution to the 1D spin density $s_z(x)$ via the
mechanism discussed in Sec.~\ref{sec:cdw}, 
the $s_z$ correlations are dominated by the $2q_F$
oscillatory part,
\[
\langle s_z(x) s_z(x') \rangle \sim \cos[2q_F(x-x')] \,
|x-x'|^{-2\eta} \;.
\]
For the distances $|x-x'|$ where the correlations are
not already vanishingly small, the cosine factor stays
essentially at unity. Hence the SDW correlations show a
pronounced {\em ferromagnetic} character.
This reasoning offers an explanation for the ferromagnetic tendencies observed
in Ref.\cite{tans}. Our explanation could be experimentally checked
by increasing the doping and hence $q_F$.
Eventually, antiferromagnetism should
be recovered for sufficiently strong doping.

\subsection{Superconductivity}

Predominant superconducting (SC) features have been predicted
recently in two-chain models despite the 
repulsive nature of the Coulomb  interaction \cite{krotov,fabrizio,schulz}.
For the nanotube, the inter-sublattice singlet and all
triplet pairing operators cause irrelevant terms and therefore play no role.
The dominant contribution then comes from the
{\em intra-sublattice singlet pairing} (SC0) operator
\[
\hat{O}_{SC0} \sim \sum_{p\alpha\sigma} \sigma \psi^{}_{p\alpha\sigma} 
\psi^{}_{p-\alpha-\sigma}\;.
\]
This gives the bosonized form 
\begin{eqnarray} \nonumber
 \hat{O}_{SC0} &\sim& \sum_{r\alpha \sigma} \,\sigma
\widetilde{\psi}^{}_{r\alpha \sigma}
\widetilde{\psi}^{}_{-r,-\alpha,-\sigma} \\ \label{osc0}
&\sim& \cos[\sqrt{\pi} \phi_{c+}]
\cos[\sqrt{\pi} \theta_{c-}] \\ \nonumber &\times&
\cos[\sqrt{\pi} \theta_{s+}] 
\sin[\sqrt{\pi} \phi_{s-}]  - (\cos \leftrightarrow \sin) \;.
\end{eqnarray}
For $T>T_b$, this operator has scaling dimension $\eta=(3+1/K)/4$
and is therefore irrelevant. In the
intermediate temperature regime $T_f<T<T_b$,  we find
$\eta=(3+2/K)/8$, and $\hat{O}_{SC0}$ is only subleading.
However, for $T<T_f$, we obtain 
\begin{equation}\label{scc}
\langle\hat{O}_{SC0}(x)\hat{O}_{SC0}(x')\rangle
\sim  |x-x'|^{-1/2K}\;.
\end{equation}
Therefore singlet superconductivity becomes the dominant instability 
at extremely low temperatures and for externally screened 
interactions with $K>1/2$.  In practice,
superconducting correlations are thus of little importance
in carbon nanotubes.
  
\subsection{Tunneling Density of States}
\label{sec:tunn}

Under typical experimental conditions,  the contact between a SWNT and 
the attached Fermi-liquid leads is not always adiabatic. For
nonadiabatic contacts, the physics of the 
conductance is related to single-electron tunneling
into the nanotube, which in turn is governed by the
tunneling density of states (TDOS). A similar situation
also arises in tunneling microscope experiments.
Here the electron tunneling occurs close to the
impurity position, the role of the impurity being
played by the probe itself. 
Therefore let us now discuss the TDOS of a SWNT.
The presence of an open boundary, either due to the finite
length of the nanotube or due to a strong impurity,  can alter
correlation functions and, in particular, modify the
anomalous exponents \cite{kf}.
Hence, in calculating the TDOS close to the end of
the SWNT, we have to employ the {\em open boundary bosonization} method.
The latter can be developed for nanotubes in full
analogy to spin chains \cite{egaff} and quantum wires \cite{fabrizio95}.
Here we shall give the main
ideas of the approach but avoid technical details,
which the interested reader can find in Refs.\cite{fabrizio95,egaff}.

For simplicity, let us consider a 
semi-infinite nanotube, $x>0$.
The left-moving electrons in the nanotube then undergo perfect
reflection at $x=0$,
where they are transformed into a right-mover going
back to $x\to \infty$.
Mathematically, the perfect reflection leads to
boundary conditions imposed on the right- and left-moving
electron field operators at $x=0$,
\begin{equation} \label{OBB:bcond}
\widetilde{\psi}_{+\alpha\sigma}(0)=\exp(i\delta_0)\,
\widetilde{\psi}_{-\alpha\sigma}(0)\;.
\end{equation}
Here $\delta_0$ is a scattering phase shift, which depends
on the shape-confining potential. For example,
if the end of the nanotube has dangling (but hydrogen-saturated) bonds, 
 $\delta_0$ is different compared to a closing cap, 
i.e., if the nanotube closes into a 
fullerene half-sphere.
However, this phase shift is of no further interest here
and can be chosen as zero
by shifting the position of the boundary \cite{fabrizio95}. 

Due to the boundary condition (\ref{OBB:bcond}),
the right- and left-movers are not independent fields anymore. 
It therefore makes sense to switch to a description in terms of only
a right-moving field which is now defined for all $x$,
\begin{equation} \label{OBB:rmfield}
\Psi_{\alpha\sigma}(x)=\left\{
\begin{array}{l}
\widetilde{\psi}_{+\alpha\sigma}(x)\quad (x>0)\;,\\
\widetilde{\psi}_{-\alpha\sigma}(-x)\quad (x<0)\;.
\end{array} \right.
\end{equation}
The advantage of this representation is that the right-moving
field operator $\Psi_{\alpha\sigma}$, being defined on the infinite interval,
can be standardly bosonized,
\begin{equation} \label{OBB:rmbos}
\Psi_{\alpha\sigma}(x)=\frac{1}{\sqrt{2\pi a}}
\exp[i\varphi_{\alpha\sigma}(x)]\;,
\end{equation}
where $\varphi_{\alpha\sigma}$ is a chiral Bose field.
For brevity, we have omitted the Klein factors and the oscillating
spatial exponent. Both cancel out in what follows.
It is then helpful to define the chiral fields
\begin{eqnarray*}
\Phi_{c+} &=& \frac{1}{4\sqrt{\pi}} \sum_{\alpha\sigma} 
\varphi_{\alpha\sigma} \;, \\
\Phi_{c-} &=& \frac{1}{4\sqrt{\pi}} \sum_{\alpha\sigma} 
\alpha \varphi_{\alpha\sigma} \;, \\
\Phi_{s+} &=& \frac{1}{4\sqrt{\pi}} \sum_{\alpha\sigma} 
\sigma \varphi_{\alpha\sigma} \;, \\
\Phi_{s-} &=& \frac{1}{4\sqrt{\pi}} \sum_{\alpha\sigma} 
\alpha\sigma  \varphi_{\alpha\sigma} \;.
\end{eqnarray*}
In terms of these fields, Eq.~(\ref{OBB:rmbos}) takes
the form
\begin{equation}
\Psi_{\alpha\sigma}=\frac{1}{\sqrt{2\pi a}}
\exp\left\{i\sqrt{\pi}\left[
\Phi_{c+}+\alpha\Phi_{c-}+\sigma\Phi_{s+}
+\alpha\sigma\Phi_{s-}
\right]\right\}\;.
\label{OBB:rmbosbis}
\end{equation}
If electron-electron interactions are neglected,
the $\Phi_{j\delta}$ are free chiral fields
with the Hamiltonian
\begin{eqnarray}
&~&~H_0=-i\sum_{r\alpha\sigma}r\int_0^\infty dx\,
\widetilde{\psi}^\dagger_{r\alpha\sigma}\partial_x
\widetilde{\psi}^{}_{r\alpha\sigma}=\nonumber\\
&~&~-i\sum_{\alpha\sigma}\int_{-\infty}^\infty dx\,
\Psi^\dagger_{\alpha\sigma}\partial_x
\Psi^{}_{\alpha\sigma}= 
v \sum_{j\delta} \int dx \, (\partial_x \Phi_{j\delta})^2 \;.
\label{OBB:Hnot}
\end{eqnarray}
In the presence of interactions, however, the nonlocal
representation (\ref{OBB:rmfield}) causes
nonlocal interaction terms of the form $\rho(x)\rho(-x)$
such that the electron densities are coupled at mirror-imaged
points. 
Nevertheless, this interaction remains quadratic in the Bose field
and can be diagonalized. Following the steps of
Ref.\cite{fabrizio95}, the appropriate Bogoliubov rotation is
\begin{equation}
\Phi_{c+}(x)\rightarrow c\,\Phi_{c+} (x) - s\,\Phi_{c+}(-x)\;,
\label{OBB:bog}
\end{equation}
where $c=\cosh\phi_0$ and $s=\sinh\phi_0$ parametrize 
the Luttinger liquid exponent $K=\exp(2\phi_0)$.

Two remarks are in order here. (1) We only consider the Luttinger
liquid phase $T>T_b$ in the following.
Then the only important interaction term
is $H^{(0)}_{\alpha\rm{FS}}$. 
In the low-temperature massive phases, the 
single-electron TDOS develops a gap, but
there could still be gapless
higher-order multi-electron scattering processes.
(2) We consider the zero-momentum limit for the charge exponent $K$,
i.e., distances (times) larger than the Coulomb screening length.
The behavior of the correlation functions for distances
closer to the boundary than the screening length is fairly 
complex, see Appendix C  of Ref.\cite{fabrizio95}.

Combining Eqs.~(\ref{OBB:rmbosbis}) and  (\ref{OBB:bog}), we obtain
the main formula of the open boundary bosonization,
\begin{eqnarray}
&~&~\Psi_{\alpha\sigma}(x)=\frac{1}{\sqrt{2\pi a}}
\exp\left\{i\sqrt{\pi}\left[
c\Phi_{c+}(x)-s\Phi_{c+}(-x)\right.\right.\nonumber\\
&~&~\left.\left.+\alpha\Phi_{c-}(x)+\sigma\Phi_{s+}(x)
+\alpha\sigma\Phi_{s-}(x)
\right]\right\}\;.
\label{OBB:bosfin}
\end{eqnarray}
It follows that close to the boundary,
i.e., for $\max (x,y)\ll vt$, the single-electron Greens
function is of the form
\begin{equation}
\langle \Psi^\dagger (x,t)\Psi(y,0)\rangle
\sim t^{-(1/K+3)/4}\;.
\label{OBB:GF}
\end{equation}
The boundary scaling dimension of the electron field
operator is therefore  
\begin{equation}
\bar{\Delta}=\frac{1}{8K}+\frac{3}{8}\;,
\label{OBB:scbound}
\end{equation}
as opposed to its bulk scaling dimension,
\begin{equation}
\Delta=\frac{1}{16}\left(\frac{1}{K}+K\right)+\frac{3}{8}\;.
\label{OBB:scbulk}
\end{equation}
Making use of the text-book definition of the density of states
as the imaginary part of the single-electron Greens function,
we obtain for the TDOS at the end of the nanotube,
\begin{equation} \label{OBB:TDOS}
\rho_{\rm end}(\omega)\sim \omega^{2\bar{\Delta}-1}\;,
\end{equation}
which is different from the bulk density of states 
\begin{equation} \label{bulk}
\rho_{\rm bulk}(\omega)\sim \omega^{2\Delta-1}\;.
\end{equation}
Since $\bar{\Delta}>1/2$ for $K<1$,
the TDOS always vanishes with the frequency (or the temperature)
approaching zero, similar
to the situation encountered in quantum wires \cite{kf}.
For a Fermi liquid, we would instead have $\Delta=\bar{\Delta}=1/2$, and
the exponents in Eqs.~(\ref{OBB:TDOS}) and
(\ref{bulk}) are both zero.
For the $(10,10)$ nanotube of Ref.\cite{tans},
taking our previous estimate $K\simeq 0.18$, the exponents in 
Eqs.~(\ref{OBB:TDOS}) and (\ref{bulk}) are
$1.13$ and $0.46$, respectively. Therefore
single-electron tunneling is 
strongly suppressed close to the end of the nanotube.
Finally, we stress again that these results are 
also valid in the vicinity of a
strong impurity. At low energy scales,
the SWNT is effectively cut into two independent parts 
because of the impurity, see Sec.~\ref{sec:trans}.

\section{Transport and Conductance Laws}
\label{sec:trans}

In this section, we discuss transport through 
a SWNT.  Since non-Fermi liquid laws are
only pronounced if contact resistances 
between the attached leads and the nanotube
are smaller than $\approx h/e^2$, we assume
good contact to the transport voltage sources.
Since there are four transport
channels ($j\delta$), a SWNT adiabatically connected
to external leads exhibits the
perfect conductance $G=G_0=4e^2/h$ in the absence
of impurity backscattering.  Even in the presence of
weak backscattering, $G$ approaches $G_0$  at sufficiently
high temperatures.
 
The effect of impurities on the conductance can be very
pronounced in a Luttinger liquid \cite{kf}.
In a SWNT, there are several possible sources for 
impurities as discussed in the Introduction. 
The dominant coupling to the impurity is due to the CDW$\pi$
operators $\hat{O}_1$ and $\hat{O}_2$ in Eqs.~(\ref{ocdw1}) and
(\ref{ocdw2}). Since both operators have essentially the
same effect, we focus on $\hat{O}_1$ and get the form
\begin{eqnarray} 
\nonumber
H_{\rm imp} & =&  \int dx\, m(x)
\{ \cos[\sqrt{\pi}\theta_{c+}+2q_F x]
 \cos[\sqrt{\pi}\theta_{c-}]\\ 
\label{impop}
&\times&
 \sin[\sqrt{\pi}\theta_{s+}]
 \sin[\sqrt{\pi}\theta_{s-}] + (\cos\leftrightarrow \sin) \} \;.
\end{eqnarray}
From the perturbation series in $m(x)$ for the partition sum, one can check
that the higher-order operators of Sec.~\ref{sec:wigner}
are implicitly contained in Eq.~(\ref{impop}).
There is no need to explicitly take them into account in
$H_{\rm imp}$.

Let us analyze the consequences of this Hamiltonian
for the simplest case of a single pointlike impurity,
$m(x)=m\delta(x)$. We assume an impurity without
 any internal dynamics, e.g., a
frozen twist of the nanotube.  
The conductance corrections $\delta G$ defined by
$G=G_0-\delta G$ 
depend on the temperature regime
under study. Perturbation theory yields in order $m^2$
the power laws
\begin{eqnarray} \label{pw1}
\delta G &\sim& T^{(K-1)/2}  \quad (T>T_b)\;,\\ \label{pw2}
 &\sim& T^{(2K-5)/4}  \quad (T_f < T < T_b) \;.
\end{eqnarray} 
At very low temperatures, $T<T_f$, one is in the strong-coupling
regime and perturbation theory does not apply anymore.
In fact, we find that the conductance is totally
suppressed at zero temperature.
In the regime $T<T_f$, the average value of the impurity operator
(\ref{impop}) vanishes, while its correlation function
decays exponentially in time.
At first sight, the impurity has therefore no effect.
However, it may still generate a relevant operator.
In order to see how this happens, we calculate the
second-order correction to the action due to the
local impurity operator, 
\begin{eqnarray*}
\delta S^2 &\sim& m^2 \int d\tau \int d\tau'
\, \exp[-\gamma|\tau-\tau'|] \\ &\times&
\cos[\sqrt{\pi}\theta_{c+}(0,\tau)]
\cos[\sqrt{\pi}\theta_{c+}(0,\tau')]\;,
\end{eqnarray*}
where $\gamma \sim m_f$ characterizes the decay rate of 
correlations of the gapped degrees of freedom  involved 
in the impurity operator (\ref{impop}).
Since we are interested in the low-energy (long-time)
behavior of the system, we are eligible to contract the
$\tau$ and $\tau'$ variables in the above integration.
Then, using the operator identity (\ref{identity}),
the total charge field satisfies the effective Hamiltonian
\begin{equation}\label{effimp}
H_{\rm eff}[\theta_{c+}]=H_0[\theta_{c+}]+
\frac{\Lambda}{\pi a}\cos[\sqrt{4\pi K}\, \theta_{c+}(0)]\;,
\end{equation}
where the effective impurity strength is estimated to be
$\Lambda \sim m^2/m_f$, and
we have rescaled $\theta_{c+}\to \sqrt{K} \theta_{c+}$. 
Of course, the exponent found in Eq.~(\ref{pw3})
below is independent of $\Lambda$.
Note that the result (\ref{effimp}) also follows from 
Eq.~(\ref{o12})  for $T<T_f$ by virtue of the pinning 
 conditions discussed in Sec.~\ref{sec:maj}.

The effective Hamiltonian (\ref{effimp})
is now identical to the one describing an impurity
in a spinless Luttinger liquid.
The reasoning of Kane and Fisher \cite{kf}
therefore fully applies and yields 
the low-temperature conductance 
\begin{equation}\label{pw3}
G\sim T^{-2+2/K} \;.
\end{equation}
This implies a total suppression of transport through
the nanotube at $T=0$ in the presence of a single,
arbitrarily weak impurity.

Finally, we mention that  Luttinger liquid 
behavior could also be observed in experiments on
crossed nanotubes.  Nonequilibrium transport
through crossed Luttinger liquids has been 
studied theoretically in Ref.\cite{komnik}.
Following this analysis,
a distinct nonlinear dependence of the current
through one SWNT ($j=1$) on the cross voltage applied to the
other SWNT ($j=2$) can be expected.  
Assuming a point-like contact between both nanotubes  at $x=0$,
the only important coupling mechanism is of 
electrostatic nature, 
\begin{equation} \label{cross}
H_c = \widetilde{\lambda} \; \rho_1(0) \rho_2(0)\;,
\end{equation}
which becomes relevant if the scaling dimension $\eta$ of the
individual local density operators $\rho_{j=1,2}(0)$ satisfies
$\eta<1/2$.  This condition can only be met by
the higher-order CDW$\pi$ contributions in Sec.\ref{sec:wigner}.
In the Luttinger liquid regime $T>T_b$, the leading
operator is $\hat{O}_{CDW\pi}^4$ in Eq.~(\ref{o4}), with
$\eta=4K$.  The coupling (\ref{cross}) is then irrelevant 
unless the condition $K<1/8$ is fulfilled.
According to Eq.~(\ref{longr}), such strong correlations
 could be achieved by studying
sufficiently long nanotubes.  Furthermore, at lower 
temperatures $T<T_b$, the dominant coupling comes
from $\hat{O}_1^2$ in Eq.~(\ref{o12}), with $\eta=K$.  
Therefore the relevancy condition
is now less stringent, $K<1/2$, and the effects predicted
in Ref.\cite{komnik} should show up already for short
nanotubes  at sufficiently low temperatures.

\section{Impurity screening}
\label{sec:imp}

In this section, we study the electronic screening cloud induced
by an impurity sitting at, say, $x=0$.   This could be
a substitutional atom replacing one of the carbon atoms
 on the tube surface (like B or N), 
an intercalating atom (like Na) sitting inside the nanotube at some
radius $r<R$, or  an incident positron. 
The (externally unscreened)
 Fourier transformed potential 
for  impurity charge $Ze$ is \cite{liu}
\begin{equation}\label{impot}
\widetilde{V}_{\rm imp}(q)=4\pi Z e^2 I_0(qr) K_0(qR) \;,
\end{equation}
with the modified Bessel functions $I_0$ and $K_0$ \cite{abramowitz}.
The small-$q$ response of the $\pi$
electrons  to this perturbation can be computed  in an exact manner
by linear screening, and the RPA treatment and
the corresponding results of Ref.\cite{liu} fully apply.
However, for the finite wavevectors listed in
Eq.~(\ref{wavelength}), linear screening  is known 
to break down in one dimension
\cite{egger95,fabrizio95,matveev,egger97b}.
The 1D Friedel oscillation cannot be computed by an RPA-like
treatment.

The impurity strength $m$
determining the
amplitude of the Friedel oscillation is given by
$\widetilde{V}_{\rm imp}(2q_F)$ [for wavelength
$\lambda=\pi/q_F$], 
or $\widetilde{V}_{\rm imp}(2k_F)$
[for $\lambda=\pi/k_F\simeq \pi/(k_F\pm q_F)$].
In the latter case, $m$ is strongly
dependent on the position of the impurity.
If the impurity sits in the center of the nanotube ($r=0$), we
get exponential suppression of 
$m$ due to Eq.~(\ref{impot}).  This implies 
a strongly reduced amplitude of the Friedel oscillation.
However, the actual value of $m$ does {\em not}\, affect the power laws
reported below.  These are universal, i.e., independent of 
strength, position $r$, or  nature of the impurity.

As discussed in Sec.~\ref{sec:cdw}, the 1D density operator
 $q(x)$ has
several contributions. 
The expectation value of the slow part $\rho(x)$ in
the presence of the impurity can be obtained
by RPA \cite{liu} and is not further discussed here.
The contributions beyond $\rho(x)$ 
are due to the CDW order operators  in Sec.~\ref{sec:susc}.
These arise from mixing the different types of 1D fermion
operators. 

Following Ref.\cite{egger95}, it is 
straightforward to extract the power law decay 
of the Friedel oscillation.  For $T>T_b$, 
the CDW$\pi$ operators $\hat{O}_1$ and
$\hat{O}_2$ specified in Eqs.~(\ref{ocdw1}) and (\ref{ocdw2})
lead to the simultaneous presence of the wavelengths
$\pi/q_F$ and $\pi/(k_F\pm q_F)$.  Contrary to the  RPA
analysis \cite{liu}, our exact treatment of
Coulomb interactions yields for $x\gg a$ \cite{egger95,foot3},
\begin{eqnarray}\label{qf2}
\langle q_{2q_F}(x)\rangle &\sim & \cos[2q_F x] \, (x/x_0)^{-(3+K)/4} 
 \quad (x\gg x_0)  \\ \nonumber
&\sim& \cos[2q_F x]\, (x/x_0)^{-(1+K)/2} \quad (x\ll x_0) \;, \\
\label{kf2} 
\langle q_{2k_F}(x) \rangle &\sim& 
\cos[2(k_F\pm q_F) x] \, (x/x_0)^{-(3+K)/4} 
\quad (x\gg x_0) \\ \nonumber
&\sim& \cos[2(k_F\pm q_F) x] \, (x/x_0)^{-(1+K)/2} \quad
(x\ll x_0) \;.
\end{eqnarray}
We omit possible phase shifts here. Furthermore,
$x_0$ sets the appropriate
length scale, which can be different
for the various oscillation periods since it
depends on the impurity strength,
$x_0\sim m^{-4/(1-K)}$. 
The asymptotic exponents for $x\gg x_0$ can
be obtained from  open boundary bosonization, see Sec.\ref{sec:tunn},
while the behavior close to the impurity follows from 
 perturbation theory in $m$.
Here the important point is that the Friedel oscillation is
always {\em slower} than the standard $1/x$ Fermi liquid  result.
Furthermore, the decay becomes slower as the impurity 
is approached.  Following the reasoning of Ref.\cite{matveev},
the slow decay of the Friedel oscillation is the physical 
reason for the conductance suppression
by a single impurity discussed in Sec.\ref{sec:trans}.
Interestingly, we have again several different
oscillation periods, namely a slow one,
 $\lambda=\pi/q_F$, which is superimposed onto rapid
oscillations with wavelength $\lambda=\pi/(k_F\pm q_F)$.
The Friedel oscillations induced by a strong impurity
or by the ends of the nanotube can be quite pronounced and should be 
detectable by placing a STM tip close to the nanotube.

The power laws (\ref{qf2}) and (\ref{kf2}) 
are modified due to the $8q_F$ operator (\ref{o4}) 
in the case of extremely strong correlations.
This operator plays no role at temperatures $T<T_b$,
but in the Luttinger liquid regime it becomes dominant as soon as
 $K<1/5$.  Instead of Eqs.~(\ref{qf2}) and (\ref{kf2}), we then find
the wavelength $\pi/4q_F$,
\begin{eqnarray}\label{8qf}
\langle q_{8q_F}(x)\rangle &\sim& \cos[8q_F x] \, (x/x_0)^{-4K}  \quad 
(x\gg x_0)  \\ \nonumber
&\sim& \cos[8q_F x] \, |\ln(x/x_0)| \quad (x\ll x_0) \;.
\end{eqnarray}
The only logarithmically
slow decay of the Friedel oscillation close to the impurity 
follows from  Ref.\cite{saleur}.

The behavior of the Friedel oscillations discussed so far
applies only to the ``Luttinger liquid'' regime, $T> T_b$. 
In the intermediate temperature regime, $T_f< T< T_b$, 
and for not too strong correlations, $1/2 <K <1$,
the exponents in these power laws become $-(3+2K)/8$
for $x\gg x_0$, and $(1-2K)/4$ for $x\ll x_0$, respectively.
For strong correlations, $K<1/2$,  the squared CDW$\pi$
operators of Sec.~\ref{sec:wigner}
are more important than $\hat{O}_1$ and $\hat{O}_2$.
In that case, only the slow oscillation period
 $\pi/2q_F$  corresponding to Eq.~(\ref{o12})  shows up, 
and we obtain
\begin{eqnarray} \label{qf4}
\langle q_{4q_F}(x) \rangle &\sim& \cos[4q_F x] \, (x/x_0)^{-K}  \quad 
(x\gg x_0)  \\ \nonumber
&\sim& \cos[4q_F x] 
\, |\ln(x/x_0)| \quad
(x\ll x_0)\;.
\end{eqnarray}
Finally, at the lowest temperatures, $T<T_f$, the same result
is found for $K<1/2$.  For $K>1/2$, 
the perturbative exponent $\sim x^{1-2K}$ 
applies to the short-distance behavior,
with the same asymptotic exponent 
as in Eq.~(\ref{qf4}).

The overall behavior of Friedel oscillations in nanotubes
is considerably more complicated than predicted by
linear screening.  
We find several different slow algebraic
 power laws  in various regions
of  parameter space. They are all slower than the standard
$1/x$ decay. Furthermore, there exist different wavelengths
for the respective dominant contributions.
 If measurements of the
charge screening cloud become feasible, one could extract
the important correlation parameter $K$ and directly observe
the various regions of the phase diagram corresponding to
Table \ref{table1} from the Friedel oscillation.

\section{Conclusions}
\label{sec:conc}

In this work we have formulated and analyzed
the effective low-energy theory
of isolated single-wall carbon nanotubes.  
Long and clean nanotubes have already been fabricated, 
and we  believe  that 
the non-Fermi liquid effects discussed here will be 
observed in the near future.  

In fact, there is some evidence
that one of these effects, namely the predominance
of ferromagnetic correlations, has already been observed 
in Ref.\cite{tans}.
Other signatures of non-Fermi-liquid behavior include
anomalous interaction-dependent power laws for the
conductance or the tunneling density of states. 
Nanotubes are promising candidates
for revealing the Luttinger liquid behavior generally
expected for one-dimensional metals. According to
our theory, at temperatures above the scale $T_b$
(where $T_b\approx 0.1$ mK for the setup of Ref.\cite{tans}),
a  nanotube behaves as a spin-$\frac12$ Luttinger liquid
with an additional flavor index.
At lower temperatures, more complicated phases characterized
by gaps in the neutral modes emerge.

  A special feature of 
doped nanotubes is the simultaneous occurrence of different
wavelengths for the various order parameter correlations.
The doping is characterized by a small wavevector $q_F\ll k_F$,
and one quite generally finds pronounced instabilities
for the long wavelengths $\lambda=\pi/q_F, \pi/2q_F$ and
$\pi/4q_F$, besides
the short wavelength $\lambda=\pi/(k_F\pm q_F)$.
Another interesting aspect of nanotubes is the
length ($L$) dependence of their properties.
For instance, considering
a SWNT with sufficiently small $L$, our continuum approach for
explaining ferromagnetic tendencies should eventually be replaced
by a Hund's rule applying to the ``molecule''. 
In our opinion, this transition from a molecule to  the ``solid-state'' 
limit  of large $L$ deserves further attention.

The physics of nanotubes is likely to reveal further 
surprises in the future.  Besides the exciting prospect of
finding Luttinger liquid behavior in a potentially very
clean way,  nanotubes might act as basic elements
for molecular electronics devices, e.g., as highly
conducting wires.  Another line of development
could employ networks of nanotubes, where novel
correlation effects can be anticipated.  We hope
that our theory will be useful in addressing and
resolving these issues.

\acknowledgements

We wish to thank C.~Dekker, M.H.~Devoret, H.~Grabert, 
C.L.~Kane and A.A.~Nersesyan for inspiring discussions, 
and acknowledge financial support 
by the EPSRC of the United Kingdom.

\begin{table}
\caption{ \label{table1}
Dominant correlations in an armchair SWNT as a function of
temperature and of the correlation parameter $K< 1$, with 
 the respective scaling
dimension $\eta$ and  wavelength $\lambda$. }
\bigskip
\begin{tabular}{l|l||l|l|l}
 $T$  &  $K$ & Type     & $\eta$    & $\lambda$      \\ \hline \hline
$T>T_b$ & $1/5<K$ & SDW$\pi$ &  $(3+K)/4$  &  $\pi/q_F, \pi/(k_F\pm q_F)$\\
 & $K<1/5$ & CDW$\pi$   & $4K$   & $\pi/4q_F$          \\  \hline
$T_f<T<T_b$ & $1/2<K$ & SDW$\pi$ & $(3+2K)/8$ & $\pi/q_F$ \\
  & $K<1/2$  &  CDW$\pi$   & $K$          & $\pi/2q_F$ \\ \hline
$T<T_f$ & $1/2<K$ & SC0  & $1/4K$ &  --- \\
  & $K<1/2$  & CDW$\pi$   & $K$  & $\pi/2q_F$ \\ 
\end{tabular}
\end{table}

\clearpage

\clearpage

\clearpage

\begin{references}
\bibitem{newfoot}
The extension to a semiconducting or insulating SWNT is only
straightforward for the undoped case, since otherwise
the gap will be opened away from the Fermi surface.
A bulk mass term, however, always opens a gap directly at
the Fermi surface.
\bibitem{nfoot2}
Here the Pauli matrices act on a separate Hilbert space connected
with the Klein factors and not on the $r\alpha\sigma$ quantum
states.
\bibitem{foot}
On half-filling ($q_F=0$), there are two additional 
contributions to the bosonized Hamiltonian not specified
here.  The first comes from ``$\alpha$US'' Umklapp processes with
$\alpha_1=\alpha_2=-\alpha_3= -\alpha_4$.
The second comes from $H^{(1)}_{\alpha{\rm FS}}$ given in Eq.~(\ref{fs1}),
which presents another contribution 
beyond Eq.~(\ref{bfs1}) on half-filling.  Both terms
cause a gap in the $(c+)$ sector and imply
insulating behavior at energy scales below the gap.
\bibitem{foot1}
In order to avoid confusion, we note that it
is a continuous symmetry which cannot be broken in 1D,
while Ising-type ordering is of course allowed.
\bibitem{foot3}
Above the thermal length scale $x_T=v/k_B T$, these
power laws are cut off by an exponential decay due
to thermal fluctuations.  Strictly speaking, the condition 
$x\gg x_0$ should therefore read $x_0 \ll x \ll x_T$.
\end{references}
\end{document}